\documentclass{preprint}

\usepackage{microtype}
\usepackage{graphicx}
\usepackage{booktabs}
\usepackage{hyperref}
\usepackage{xspace}

\usepackage{amsmath}
\usepackage{amssymb}
\usepackage{mathtools}
\usepackage{amsthm}

\usepackage{multirow}
\usepackage{threeparttable}
\usepackage{caption}
\usepackage{listings}
\usepackage{xcolor}

\usepackage{enumitem}
\usepackage[most]{tcolorbox}
\usepackage[capitalize,noabbrev]{cleveref}
\usepackage[textsize=tiny]{todonotes}
\newcommand{\model}{TradingAgents\xspace}
\definecolor{tauric}{HTML}{14C290}

\title{\model: Multi-Agents LLM Financial Trading Framework}

\author
{Albert Einstein~$^{1, 2}$\footnote{Correspondence E-mail: ae@mit.edu},  Nikola Tesla~$^{3}$, Isaac Newton~$^{4}$, Daniela Rus~$^{1}$\\
\vspace{1em} % Space between authors and afilliations
\normalfont{\small $^{1}$Massachusetts Institute of Technology (MIT)}\\
\normalfont{\small $^{2}$Princeton University}\\
\normalfont{\small $^{3}$Graz University of Technology}\\
\normalfont{\small $^{4}$University of Cambridge} \vspace{2em}
}

\author
{Yijia Xiao~$^{1, 3}$,  Edward Sun~$^{1, 3}$, Di Luo~$^{1, 2}$, Wei Wang~$^{1, 3}$\\
\vspace{1em} % Space between authors and afilliations
\normalfont{\small $^{1}$University of California, Los Angeles (UCLA)}\\
\normalfont{\small $^{2}$Massachusetts Institute of Technology (MIT)}\\
\normalfont{\small $^{3}$Tauric Research\footnote{Tauric Research Organization: https://tauric.ai}} \vspace{2em}
}

\begin{document}

\maketitle
\thispagestyle{firstpagestyle} % Draws the header on the first page

\begin{abstract}
Significant progress has been made in automated problem-solving using societies of agents powered by large language models (LLMs). In finance, efforts have largely focused on single-agent systems handling specific tasks or multi-agent frameworks independently gathering data. However, the multi-agent systems' potential to replicate real-world trading firms' collaborative dynamics remains underexplored. \texttt{TradingAgents} proposes a novel stock trading framework inspired by trading firms, featuring LLM-powered agents in specialized roles such as fundamental analysts, sentiment analysts, technical analysts, and traders with varied risk profiles. The framework includes Bull and Bear researcher agents assessing market conditions, a risk management team monitoring exposure, and traders synthesizing insights from debates and historical data to make informed decisions. By simulating a dynamic, collaborative trading environment, this framework aims to improve trading performance. Detailed architecture and extensive experiments reveal its superiority over baseline models, with notable improvements in cumulative returns, Sharpe ratio, and maximum drawdown, highlighting the potential of multi-agent LLM frameworks in financial trading. \textsc{TradingAgents} is available at \url{https://github.com/TauricResearch/TradingAgents}.
\end{abstract}

\section{Introduction}
Autonomous agents leveraging Large Language Models (LLMs) present a transformative approach to decision-making by replicating human processes and workflows across various applications. These systems enhance the problem-solving capabilities of language agents by equipping them with tools and enabling collaboration with other agents, effectively breaking down complex problems into manageable components \citep{park2023generativeagentsinteractivesimulacra, havrilla2024teachinglargelanguagemodels, talebirad2023multiagentcollaborationharnessingpower, tang2024medagentslargelanguagemodels}. One prominent application of these autonomous frameworks is in the financial market—a highly complex system influenced by numerous factors, including company fundamentals, market sentiment, technical indicators, and macroeconomic events.

Traditional algorithmic trading systems often rely on quantitative models that struggle to fully capture the complex interplay of diverse factors. In contrast, LLMs excel at processing and understanding natural language data, making them particularly effective for tasks that require textual comprehension, such as analyzing news articles, financial reports, and social media sentiment. Additionally, deep learning-based trading systems often suffer from low explainability, as they rely on hidden features that drive decision-making but are difficult to interpret. Recent advancements in multi-agent LLM frameworks for finance have shown significant promise in addressing these challenges. These frameworks create explainable AI systems, where decisions are supported by evidence and transparent reasoning \citep{li2023tradinggpt, wang2024quantagentseekingholygrail, yu2024fincon}, demonstrating the potential in financial applications.

Despite their potential, most current applications of language agents in the financial and trading sectors face two significant limitations:

\textbf{Lack of Realistic Organizational Modeling:} Many frameworks fail to capture the complex interactions between agents that mimic the structure of real-world trading firms \citep{li2023tradinggpt, wang2024quantagentseekingholygrail, yu2024fincon}. Instead, they focus narrowly on specific task performance, often disconnected from the organizational workflows and established human operating procedures proven effective in trading. This limits their ability to fully replicate and benefit from real-world trading practices.

\textbf{Inefficient Communication Interfaces:} Most existing systems use natural language as the primary communication medium, typically relying on message histories or an unstructured pool of information for decision-making \citep{park2023generativeagentsinteractivesimulacra, qian2024chatdevcommunicativeagentssoftware}. This approach often results in a ``telephone effect'', where details are lost, and states become corrupted as conversations lengthen. Agents struggle to maintain context and track extended histories while filtering out irrelevant information from previous decision steps, diminishing their effectiveness in handling complex, dynamic tasks. Additionally, the unstructured pool-of-information approach lacks clear instructions, forcing logical communication and information exchange between agents to depend solely on retrieval, which disrupts the relational integrity of the data.

In this work, we address these key limitations of existing models by introducing a system that overcomes these challenges. First, our framework bridges the gap by simulating the multi-agent decision-making processes typical of professional trading teams. It incorporates specialized agents tailored to distinct aspects of trading, inspired by the organizational structure of real-world trading firms. These agents include fundamental analysts, sentiment/news analysts, technical analysts, and traders with diverse risk profiles. Bullish and bearish debaters evaluate market conditions to provide balanced recommendations, while a risk management team ensures that exposures remain within acceptable limits. Second, to enhance communication, our framework combines structured outputs for control, clarity, and reasoning with natural language dialogue to facilitate effective debate and collaboration among agents. This hybrid approach ensures both precision and flexibility in decision-making.

We validate our framework through experiments on historical financial data, comparing its performance against multiple baselines. Comprehensive evaluation metrics, including cumulative return, Sharpe ratio, and maximum drawdown, are employed to assess its overall effectiveness.
\section{Related Work}

\subsection{LLMs as Financial Assistants}

Large Language Models (LLMs) are applied in finance by fine-tuning on financial data or training on financial corpora. This improves the model’s understanding of financial terminology and data, enabling a specialized assistant for analytical support, insights, and information retrieval, rather than trade execution.

\textbf{Fine-Tuned LLMs for Finance}

Fine-tuning enhances domain-specific performance. Examples include PIXIU (FinMA) \citep{xie2023pixiu}, which fine-tuned LLaMA on 136K finance-related instructions; FinGPT \citep{touvron2023llama2openfoundation, yang2023fingpt}, which used LoRA to fine-tune models like LLaMA and ChatGLM with about 50K finance-specific samples; and Instruct-FinGPT \citep{zhang2023instructfingpt}, fine-tuned on 10K instruction samples from financial sentiment analysis datasets. These models outperform their base versions and other open-source LLMs like BLOOM and OPT \citep{zhang2022opt} in finance classification tasks, even surpassing BloombergGPT \citep{wu2023bloomberggpt} in several evaluations. However, in generative tasks, they perform similarly or slightly worse than powerful general-purpose models like GPT-4, indicating a need for more high-quality, domain-specific datasets.

\textbf{Finance LLMs Trained from Scratch}

Training LLMs from scratch on finance-specific corpora aims for better domain adaptation. Models like BloombergGPT \citep{wu2023bloomberggpt}, XuanYuan 2.0 \citep{zhang2023xuanyuan}, and Fin-T5 \citep{lu2023bbtfin} combine public datasets with finance-specific data during pretraining. BloombergGPT, for instance, was trained on both general and financial text, with proprietary Bloomberg data enhancing its performance on finance benchmarks. These models outperform general-purpose counterparts like BLOOM-176B and T5 in tasks such as market sentiment classification and summarization. While they may not match larger closed-source models like GPT-3 or PaLM \citep{chowdhery2022palm}, they offer competitive performance among similar-sized open-source models without compromising general language understanding.

In summary, finance-specific LLMs developed through fine-tuning or training from scratch show significant improvements in domain-specific tasks, underscoring the importance of domain adaptation and the potential for further enhancements with high-quality finance-specific datasets.

\begin{figure*}[htbp]
\centerline{\includegraphics[width=\linewidth]{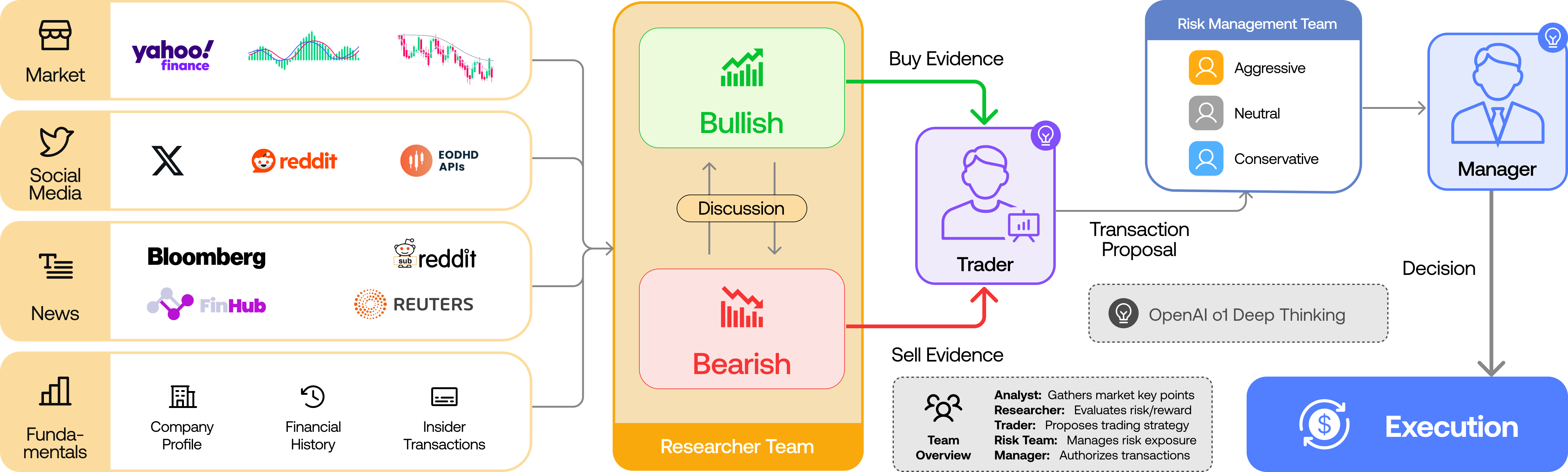}}
\caption{\textbf{\textcolor{brown}{\model}} Overall Framework Organization. \textsc{\textbf{I. Analysts Team}}: Four analysts concurrently gather relevant market information. \textsc{\textbf{II. Research Team}}: The team discusses and evaluates the collected data. \textsc{\textbf{III. Trader}}: Based on the researchers' analysis, the trader makes the trading decision. \textsc{\textbf{IV. Risk Management Team}}: Risk guardians assess the decision against current market conditions to mitigate risks. \textsc{\textbf{V. Fund Manager}}: The fund manager approves and executes the trade.}
\label{schema}
\end{figure*}

\subsection{LLMs as Traders}

LLMs act as trader agents making direct trading decisions by analyzing external data like news, financial reports, and stock prices. Proposed architectures include news-driven, reasoning-driven, and reinforcement learning (RL)-driven agents.

\textbf{News-Driven Agents}

News-driven architectures integrate stock news and macroeconomic updates into LLM prompts to predict stock price movements. Studies evaluating both closed-source models (e.g., GPT-3.5, GPT-4) and open-source LLMs (e.g., Qwen \citep{bai2023qwentechnicalreport}, Baichuan \citep{yang2023baichuan2openlargescale}) in financial sentiment analysis have shown the effectiveness of simple long-short strategies based on sentiment scores \citep{lopezlira2023chatgptforecaststockprice}. Further research on fine-tuned LLMs like FinGPT and OPT demonstrates improved performance through domain-specific alignment \citep{unveiling, sentitrade}. Advanced methods involve summarizing news data and reasoning about their relationship with stock prices \citep{beatunveiling, wang2024llmfactorextractingprofitablefactors}.

\textbf{Reasoning-Driven Agents}

Reasoning-driven agents enhance trading decisions through mechanisms like reflection and debate. Reflection-driven agents, such as FinMem \citep{finmem} and FinAgent \citep{multimodalfinmem}, use layered memorization and multimodal data to summarize inputs into memories, inform decisions, and incorporate technical indicators, achieving superior backtest performance while mitigating hallucinations \citep{ji2023mitigatinghallucinationlargelanguage}. Debate-driven agents, like those in heterogeneous frameworks \citep{xing2024designingheterogeneousllmagents} and TradingGPT \citep{li2023tradinggptmultiagentlayeredmemory}, enhance reasoning and factual validity by employing LLM debates among agents with different roles, improving sentiment classification and increasing robustness in trading decisions.

\textbf{Reinforcement Learning-Driven Agents}

Reinforcement learning methods align LLM outputs with expected behaviors, using backtesting as rewards. SEP \citep{Koa_2024} employs RL with memorization and reflection to refine LLM predictions based on market history. Classical RL methods are also used in trading frameworks that integrate LLM-generated embeddings with stock features, trained via algorithms like Proximal Policy Optimization (PPO) \citep{ding2023integratingstockfeaturesglobal, ppo}.

\subsection{LLMs as Alpha Miners}

LLMs are also used to generate alpha factors instead of making direct trading decisions. QuantAgent \citep{wang2023alpha} demonstrates this by leveraging LLMs to produce alpha factors through an inner-loop and outer-loop architecture. In the inner loop, a writer agent generates a script from a trader's idea, while a judge agent provides feedback. In the outer loop, the code is tested in the real market, and trading results enhance the judge agent. This approach enables progressive approximation of optimal behavior.

Subsequent research, such as AlphaGPT \citep{wang2023alpha}, proposes a human-in-the-loop framework for alpha mining with a similar architecture. Both studies showcase the effectiveness of LLM-powered alpha mining systems, highlighting their potential in automating and accelerating the development of trading strategies by generating and refining alpha factors.

\section{\model: Role Specialization}

Assigning LLM agents clear, well-defined roles with specific goals enables the breakdown of complex objectives into smaller, manageable subtasks. Financial trading is a prime example of such complexity, demanding the integration of diverse signals, inputs, and specialized expertise. In the real world, this approach to managing complexity is demonstrated by trading firms that rely on expert teams to collaborate and make high-stakes decisions, underscoring the multifaceted nature of the task.

In a typical trading firm, vast amounts of data are collected, including financial metrics, price movements, trading volumes, historical performance, economic indicators, and news sentiment. This data is then analyzed by quantitative experts (quants), including mathematicians, data scientists, and engineers, using advanced tools and algorithms to identify trends and predict market movements.

Inspired by this organizational structure, \model defines seven distinct agent roles within a simulated trading firm: Fundamentals Analyst, Sentiment Analyst, News Analyst, Technical Analyst, Researcher, Trader, and Risk Manager. Each agent is assigned a specific name, role, goal, and set of constraints, alongside predefined context, skills, and tools tailored to their function. For example, a Sentiment Analyst is equipped with tools like web search engines, Reddit search APIs, X/Twitter search tools, and sentiment score calculation algorithms, while a Technical Analyst can execute code, calculate technical indicators, and analyze trading patterns. More specifically, \model assumes the following teams.

\subsection{Analyst Team}

\begin{figure}[htbp]
\centerline{\includegraphics[width=0.9\linewidth]{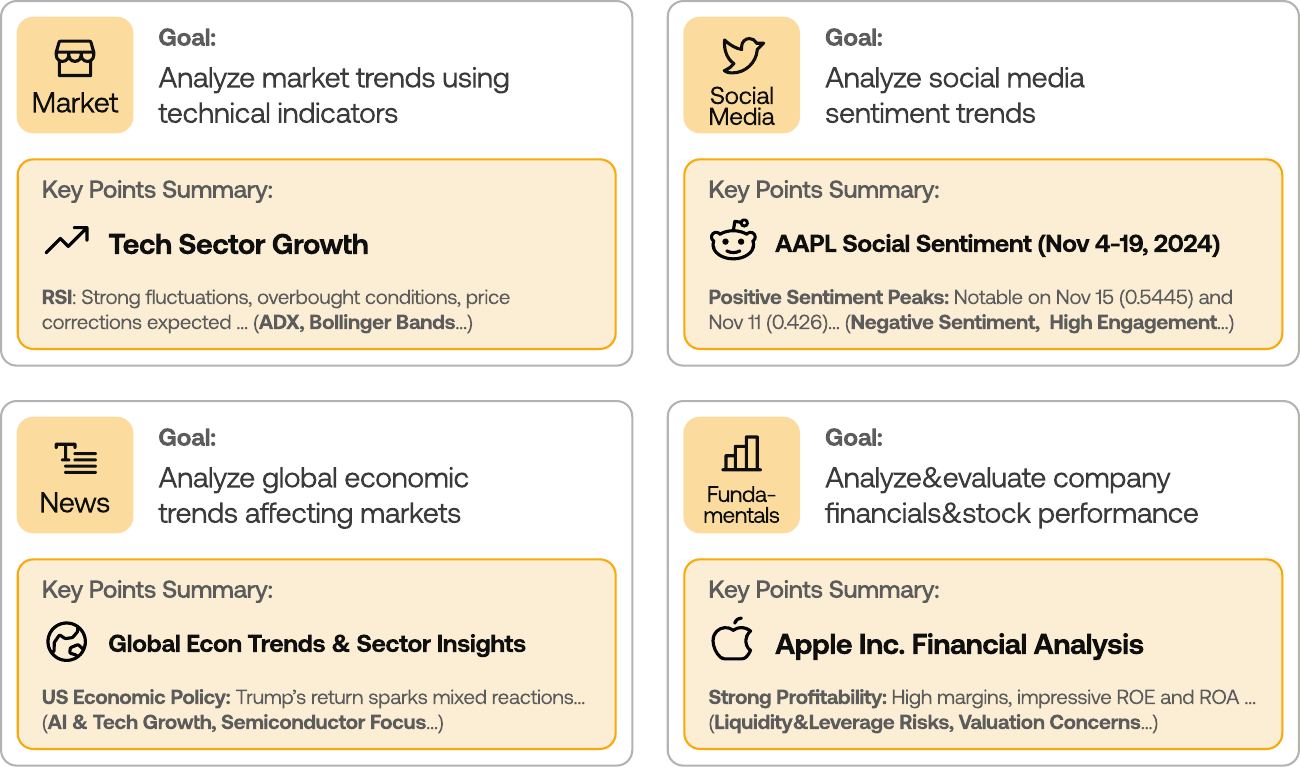}}
\caption{\model Analyst Team}

\label{fig:analyst}
\end{figure}

The Analyst Team (Figure \ref{fig:analyst}) is composed of specialized agents responsible for gathering and analyzing various types of market data to inform trading decisions. Each agent focuses on a specific aspect of market analysis, bringing together a comprehensive view of the market's conditions.

\begin{itemize}
    \item \textbf{Fundamental Analyst Agents}: These agents evaluate company fundamentals by analyzing financial statements, earnings reports, insider transactions, and other pertinent data. They assess a company's intrinsic value to identify undervalued or overvalued stocks, providing insights into long-term investment potential.
    \item \textbf{Sentiment Analyst Agents}: These agents process large volumes of social media posts, sentiment scores, and insider sentiments derived from public information and social media activity. They gauge market sentiment to predict how collective investor behavior might impact stock prices in the short term.
    \item \textbf{News Analyst Agents}: These agents analyze news articles, government announcements, and other macroeconomic indicators to assess the market's macroeconomic state, major world events, and significant company changes. They identify news events that could influence market movements, helping to anticipate sudden shifts in market dynamics.
    \item \textbf{Technical Analyst Agents}: These agents calculate and select relevant technical indicators, such as Moving Average Convergence Divergence (MACD) and Relative Strength Index (RSI), customized for specific assets. They analyze price patterns and trading volumes to forecast future price movements, assisting in timing entry and exit points.
\end{itemize}

Collectively, the Analyst Team synthesizes data from multiple sources to provide a holistic market analysis. Their combined insights form the foundational input for the Researcher Team, ensuring that all facets of the market are considered in subsequent decision-making processes.

\subsection{Researcher Team}

\begin{figure}[htbp]
\centerline{\includegraphics[width=0.9\linewidth]{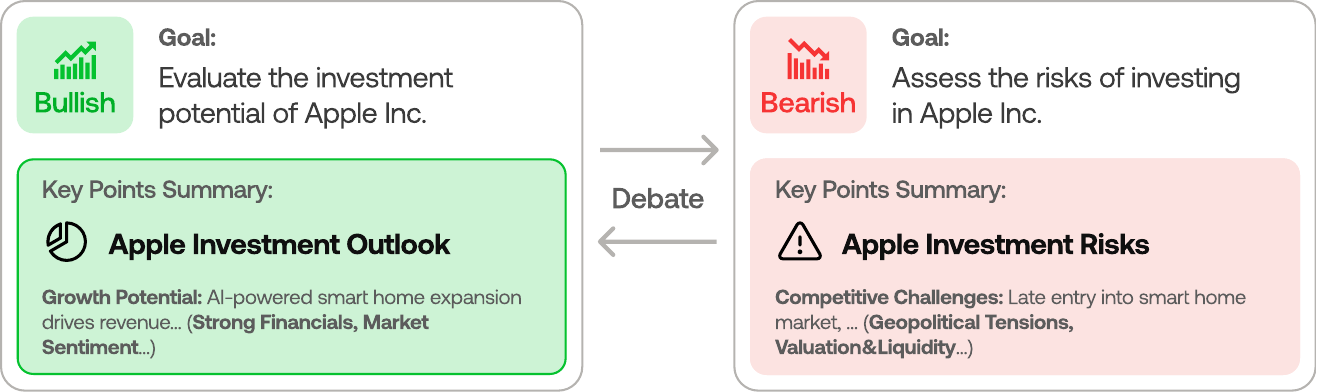}}
\caption{\model Researcher Team: Bullish Perspectives and Bearish Perspectives}
\label{fig:researcher}
\end{figure}

The Researcher Team (Figure \ref{fig:researcher}) is responsible for critically evaluating the information provided by the Analyst Team. Comprised of agents adopting both bullish and bearish perspectives, they engage in multiple rounds of debate to assess the potential risks and benefits of investment decisions.

\begin{itemize}
    \item \textbf{Bullish Researchers}: These agents advocate for investment opportunities by highlighting positive indicators, growth potential, and favorable market conditions. They construct arguments supporting the initiation or continuation of positions in certain assets.
    \item \textbf{Bearish Researchers}: Conversely, these agents focus on potential downsides, risks, and unfavorable market signals. They provide cautionary insights, questioning the viability of investment strategies and highlighting possible negative outcomes.
\end{itemize}

Through this dialectical process, the Researcher Team aims to reach a balanced understanding of the market situation. Their thorough analysis helps in identifying the most promising investment strategies while anticipating possible challenges, thus aiding the Trader Agents in making informed decisions.

\subsection{Trader Agents}

\begin{figure}[htbp]
\centerline{\includegraphics[width=0.9\linewidth]{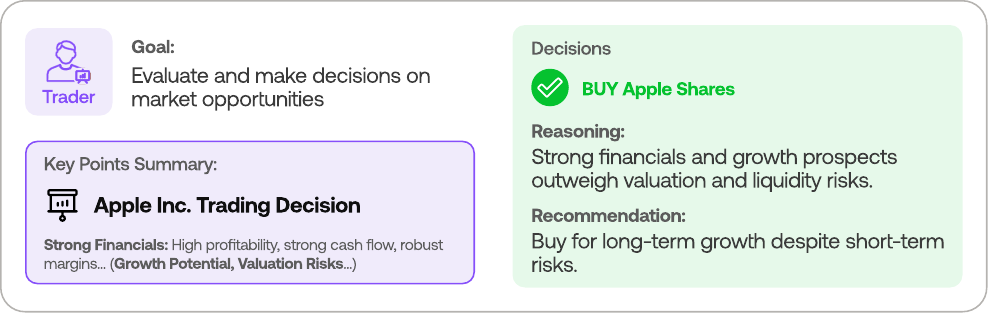}}
\caption{\model's Trader Decision-Making Process}
\label{fig:trader}
\end{figure}

Trader Agents (Figure \ref{fig:trader}) are responsible for executing trading decisions based on the comprehensive analysis provided by the Analyst Team and the nuanced perspectives from the Researcher Team. They assess the synthesized information, considering both quantitative data and qualitative insights, to determine optimal trading actions.

The tasks of \model Trader include:

\begin{itemize}
    \item Evaluating recommendations and insights from analysts and researchers.
    \item Deciding on the timing and size of trades to maximize trading returns.
    \item Placing buy or sell orders in the market.
    \item Adjusting portfolio allocations in response to market changes and new information.
\end{itemize}

Trader Agents must balance potential returns against associated risks, making timely decisions in a dynamic market environment. Their actions directly impact the firm's performance, necessitating a high level of precision and strategic thinking.

\subsection{Risk Management Team}

\begin{figure}[htbp]
\centerline{\includegraphics[width=0.9\linewidth]{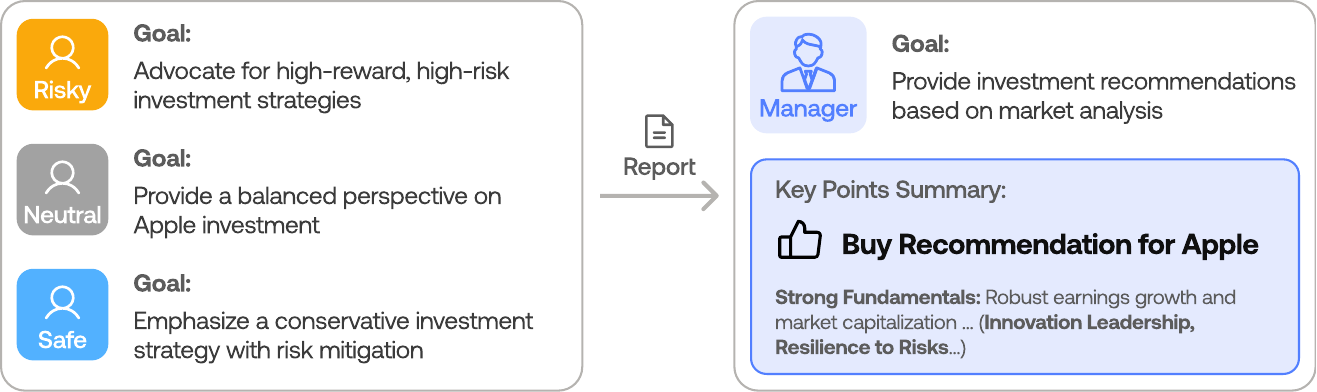}}
\caption{\model Risk Management Team and Fund Manager Approval Workflow}
\label{fig:risk_mgmt}
\end{figure}

The Risk Management Team (Figure \ref{fig:risk_mgmt}) monitors and controls the firm's exposure to various market risks. These agents continuously evaluate the portfolio's risk profile, ensuring that trading activities remain within predefined risk parameters and comply with regulatory requirements.

The responsibilities of Risk Management Team include:

\begin{itemize}
    \item Assessing factors such as market volatility, liquidity, and counterparty risks.
    \item Implementing risk mitigation strategies, such as setting stop-loss orders or diversifying holdings.
    \item Providing feedback to Trader Agents on risk exposures and suggesting adjustments to trading strategies.
    \item Ensuring that the overall portfolio aligns with the firm's risk tolerance and investment objectives.
\end{itemize}

By offering oversight and guidance, the Risk Management Team helps maintain the firm's financial stability and protect against adverse market events. They play a crucial role in safeguarding assets and ensuring sustainable long-term performance.

All agents in \model follow the ReAct prompting framework \citep{yao2023reactsynergizingreasoningacting}, which synergizes reasoning and acting. The environment state is shared and monitored by the agents, enabling them to take context-appropriate actions such as conducting research, executing trades, engaging in debates, or managing risks. This design ensures a collaborative, dynamic decision-making process reflective of real-world trading systems.

\section{\model: Agent Workflow}

\subsection{Communication Protocol}

Most existing LLM-based agent frameworks use natural language as the primary communication interface, typically through structured message histories or collections of agent-generated messages \citep{fatouros2024largelanguagemodelsbeat, li2023tradinggpt, yang2024finrobotopensourceaiagent, yang2023autogptonlinedecisionmaking}. However, relying solely on natural language often proves insufficient for solving complex, long-term tasks that require extensive planning horizons. In such cases, pure natural language communication can resemble a game of telephone—over multiple iterations, initial information may be forgotten or distorted due to context length limitations and an overload of text that obscures critical earlier details \citep{hong2024metagptmetaprogrammingmultiagent}. To address this limitation, we draw inspiration from frameworks like MetaGPT, which adopt a structured approach to communication. Our model introduces a structured communication protocol to govern agent interactions. By clearly defining each agent's state, we ensure that each role only extracts or queries the necessary information, processes it, and returns a completed report. This streamlined approach reduces unnecessary steps, lowers the risk of message corruption, and keeps interactions focused and efficient, even in complex, long-horizon tasks.

\subsection{Types of Agent Interactions}

In contrast to previous multi-agent trading frameworks, which rely heavily on natural language dialogue, \model agents communicate primarily through structured documents and diagrams. These documents encapsulate the agents' insights in concise, well-organized reports that preserve essential content while avoiding irrelevant information. By utilizing structured reports, agents can query necessary details directly from the global state, eliminating the need for lengthy conversations that risk diluting information, extending the message state indefinitely, and causing data loss. The types of documents and the information they contain are detailed below:

% \begin{itemize}
%     \setlength\itemsep{0em}
%     \setlength\leftmargin{0em}
%     \item
\textbf{I. Analyst Team}: Fundamental, sentiment, news, and technical analysts compile their research and findings into concise analysis reports specific to their areas of expertise. These reports include key metrics, insights, and recommendations based on their specialized analyses.

\textbf{II. Traders}: Traders review and analyze the reports from the analysts, carefully deliberating to produce clear decision signals. They accompany these decisions with detailed reports explaining their rationale and supporting evidence, which are later utilized by the risk management team.
% \end{itemize}

Agents engage in natural language dialogue exclusively during agent-to-agent conversations and debates. These concise, focused discussions have been shown to promote deeper reasoning and integrate diverse perspectives, enabling more balanced decisions in complex, long-horizon scenarios—a method particularly relevant to the intricate environment of trading \citep{du2023improvingfactualityreasoninglanguage}. This approach seamlessly integrates with our structured framework, as the conversation state is recorded as a structured entry within the overall agent state. The types of communication in these scenarios are detailed below:

% \begin{itemize}
%     \item
\textbf{III. Researcher Team}: Each researcher agent queries the global agent state for analyst reports and carefully forms their opinion. Two researchers represent opposing perspectives: one bullish and one bearish. They engage in natural language dialogue for $n$ rounds, as determined by the debate facilitator agent. At the conclusion, the facilitator reviews the debate history, selects the prevailing perspective, and records it as a structured entry in the communication protocol.

\textbf{IV. Risk Management Team}: The risk management team, similar to the researcher team, queries the trader's decision and accompanying report. They then deliberate from three perspectives—risk-seeking, neutral, and risk-conservative—to adjust the trading plan within risk constraints. They engage in $n$ rounds of natural language discussion, guided by a facilitator agent.

\textbf{V. Fund manager}: The fund manager reviews the discussion from the risk management team, determines the appropriate risk adjustments, and updates the trader's decision and report states within the communication protocol.
% \end{itemize}

\subsection{Backbone LLMs}

To meet the diverse complexity and speed demands of tasks in our framework, we strategically select Large Language Models (LLMs) based on their strengths. Quick-thinking models, such as \verb|gpt-4o-mini| and \verb|gpt-4o|, efficiently handle fast, low-depth tasks like summarization, data retrieval, and converting tabular data to text \citep{openai2024gpt4technicalreport}. In contrast, deep-thinking models like \verb|o1-preview| excel in reasoning-intensive tasks such as decision-making, evidence-based report writing, and data analysis. These models leverage their architectures for multi-round reasoning, producing logically sound, in-depth insights \citep{zhong2024evaluationopenaio1opportunities, wang2024planningabilitiesopenaiso1, openai2024o1}. Additionally, we prioritize models with proven reliability and scalability to ensure optimal performance across various market conditions. We also employ auxiliary expert models for specialized tasks like sentiment analysis.

Specifically, all analyst nodes rely on deep-thinking models to ensure robust analysis, while quick-thinking models handle data retrieval from APIs and tools for efficiency. Researchers and traders use deep-thinking models to generate valuable insights and support well-informed decisions. By aligning the choice of LLMs with the specific requirements of each task, our framework achieves a balance between efficiency and depth of reasoning, which is crucial for effective trading strategies.

This implementation strategy ensures that \model can be deployed without requiring a GPU, relying only on API credits. It also introduces seamless exchangeability of backbone models, enabling researchers to effortlessly replace the model with any locally hosted or API-accessible alternatives in the future. This adaptability supports the integration of improved reasoning models or finance-tuned models customized for specific tasks. As a result, \model is highly scalable and future-proof, offering flexibility to accommodate any backbone model for any of its agents.

\section{Experiments}

\definecolor{arr_green}{rgb}{0.0, 0.8, 0.0}
\definecolor{arr_red}{rgb}{1.0, 0.0, 0.0}

\subsection{Simulation Setup}
We evaluate our \model framework using a comprehensive backtesting simulation from January 1st to March 29th, 2024, across major technology stocks including Apple, Nvidia, Microsoft, Meta, and Google. \model facilitates seamless plug-and-play strategies during the simulation, enabling straightforward comparisons with any baseline. Agents make decisions based solely on data available up to each trading day, ensuring no future data is used (eliminating look-ahead bias). Based on their analysis, \model generates trading signals to buy, sell, or hold assets, which are then executed. Afterward, analysis metrics are calculated before proceeding to the next day's data.

We benchmark against five established strategies: Buy and Hold, MACD, KDJ+RSI, ZMR, and SMA (baseline descriptions in Appendix~\ref{appendix:baselines}). Performance is evaluated using four key metrics: Cumulative Return (CR), Annualized Return (AR), Sharpe Ratio (SR), and Maximum Drawdown (MDD) (formulations in Appendix~\ref{appendix:eval_metrics}).

\subsection{Back Trading}
To simulate a realistic trading environment, we utilize a multi-asset and multi-modal financial dataset comprising of various stocks such as Apple, Nvidia, Microsoft, Meta, Google, and more. Our multi-modal dataset integrates historical stock prices, news articles, social media sentiment, insider transactions, financial statements, and 60 technical indicators per asset. The dataset includes:

\textbf{Historical Stock Prices}: Open, high, low, close, volume, and adjusted close prices from January 1st, 2024, to March 29th, 2024.

\textbf{News Articles}: Daily news updates are gathered from diverse sources such as Bloomberg, Yahoo, EODHD, FinnHub, and Reddit, covering specific company developments, global events, macroeconomic trends, and government updates.

\textbf{Social Media Posts and Sentiment}: Posts from Reddit, X/Twitter, and other platforms along with sentiment scores of posts calculated by auxiliary language models.

\textbf{Insider Sentiments and Transactions}: Sentiment derived from public information, including transactions from SEDI and relevant company filings.

\textbf{Financial Statements and Earnings Reports}: Quarterly and annual reports filed by companies.

\textbf{Company Profiles and Financial History}: Descriptions of company profiles, target industries, and financial history reported by third parties.

\textbf{Technical Indicators}: Sixty standard technical analysis indicators calculated for each asset, including MACD, RSI, Bollinger Bands, etc.

% \vskip -0.2in
\begin{figure}[ht]
% \vskip 0.2in
\begin{center}
\centerline{\includegraphics[width=0.95\columnwidth]{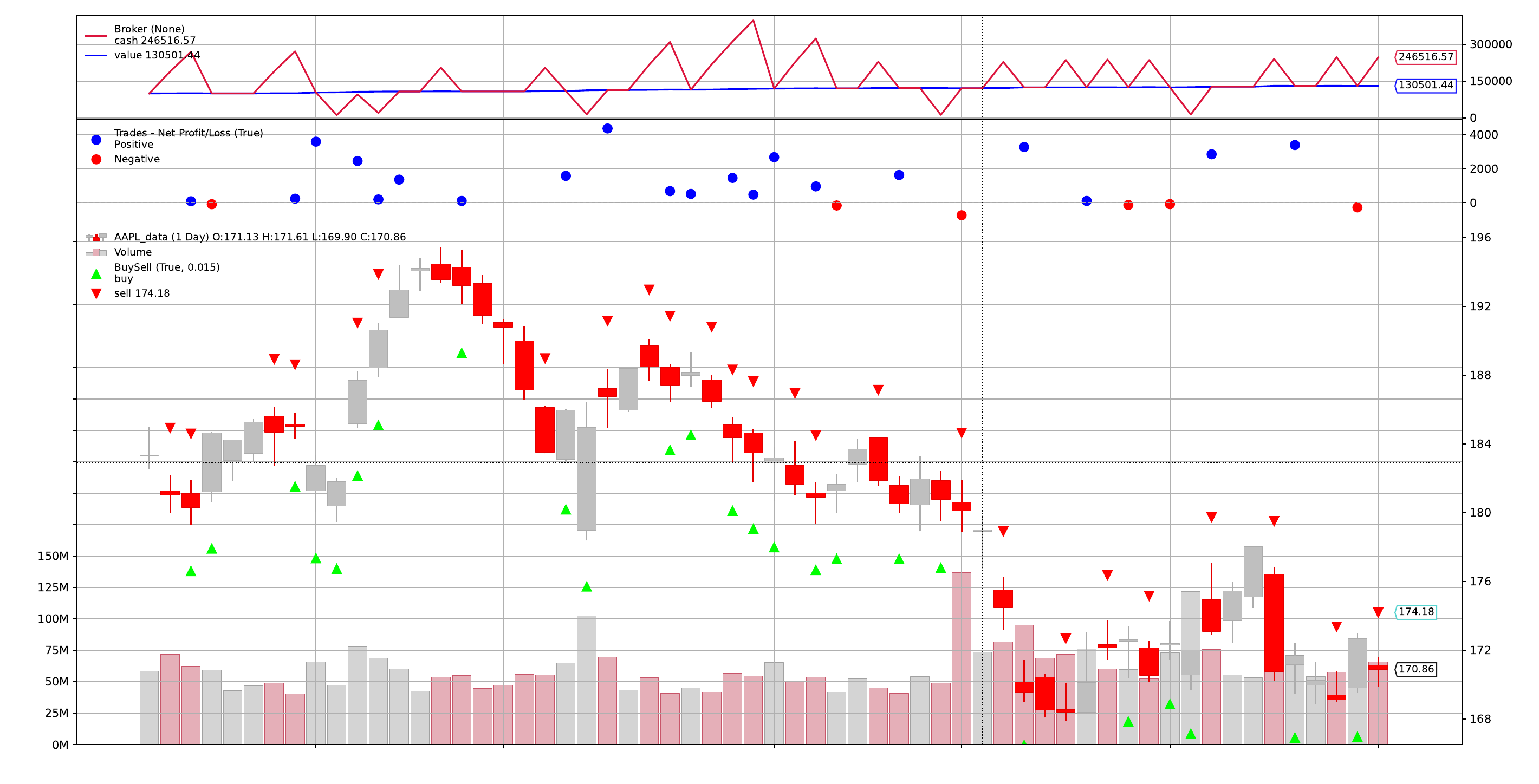}}
\caption{\textbf{\textcolor{brown}{\model}} Detailed Transaction History for \$AAPL. 
\textcolor{arr_green}{\texttt{Green}} \textcolor{brown}{\textbf{/}} \textcolor{arr_red}{\texttt{Red}} Arrows indicate \textcolor{arr_green}{\texttt{Long}} \textcolor{brown}{\textbf{/}} \textcolor{arr_red}{\texttt{Short}} Positions respectively, showing the model's trading decisions over time.}
\label{fig:aapl-transactions}
\end{center}
\vskip -0.2in
\end{figure}

\section{Results and Analysis}

In this section, we present the experimental results and analyze the performance of \model in comparison to the baseline models.

\definecolor{FBest}{RGB}{4,130,53}

\begin{table*}[htbp]
\renewcommand{\arraystretch}{0.9}
\footnotesize
\caption{Performance comparison across all methods using four evaluation metrics. Results highlighted in \textbf{\textcolor{FBest}{green}} represent the best-performing statistic for each model. The improvement row illustrates \model's performance gains over the top-performing baselines.}

\setlength{\abovecaptionskip}{0.1cm}
\centering
\begin{threeparttable}
\setlength{\tabcolsep}{2.5pt}
\resizebox{\linewidth}{!}{ 
\begin{tabular}{clcccclcccclcccc}
\toprule
\multirow{3}{*}{Categories} & \multirow{3}{*}{Models} & \multicolumn{4}{c}{AAPL} & & \multicolumn{4}{c}{GOOGL} & & \multicolumn{4}{c}{AMZN} \\ 
\cmidrule{3-6}\cmidrule{8-11}\cmidrule{13-16}
&& CR\%$\uparrow$ & ARR\%$\uparrow$ & SR$\uparrow$ & MDD\%$\downarrow$ & & CR\%$\uparrow$ & ARR\%$\uparrow$ & SR$\uparrow$ & MDD\%$\downarrow$ & & CR\%$\uparrow$ & ARR\%$\uparrow$ & SR$\uparrow$ & MDD\%$\downarrow$ \\ 
\midrule
Market 
& B\&H& -5.23& -5.09 & -1.29 & 11.90 && 7.78 & 8.09 & 1.35 & 13.04  && 17.1 & 17.6 & 3.53 & 3.80\\
\midrule
\multirow{3}{*}{\begin{tabular}[c]{@{}c@{}}Rule-based\end{tabular}}
& MACD& -1.49 & -1.48 & -0.81 & 4.53 && 6.20 & 6.26 & 2.31 & \textbf{\textcolor{FBest}{1.22}}  && - & - & - & -\\
& KDJ\&RSI& 2.05 & 2.07 & 1.64 & 1.09 && 0.4 & 0.4 & 0.02 & 1.58  && -0.77 & -0.76 &-2.25 & 1.08\\
& ZMR& 0.57 & 0.57 & 0.17 & \textbf{\textcolor{FBest}{0.86}} && -0.58 & 0.58 & 2.12 & 2.34  && -0.77 & -0.77 & -2.45 & \textbf{\textcolor{FBest}{0.82}}\\
& SMA & -3.2 & -2.97 & -1.72 & 3.67 && 6.23 & 6.43 & 2.12 & 2.34 && 11.01 & 11.6 & 2.22 & 3.97\\
\midrule
\multirow{1}{*}{\begin{tabular}[c]{@{}c@{}} Ours\end{tabular}}
& \textbf{\texttt{\model}}& \textbf{\textcolor{FBest}{26.62}} & \textbf{\textcolor{FBest}{30.5}} & \textbf{\textcolor{FBest}{8.21}} & 0.91 && \textbf{\textcolor{FBest}{24.36}} & \textbf{\textcolor{FBest}{27.58}} & \textbf{\textcolor{FBest}{6.39}} & 1.69  && \textbf{\textcolor{FBest}{23.21}} & \textbf{\textcolor{FBest}{24.90}} & \textbf{\textcolor{FBest}{5.60}} & 2.11\\
\midrule
\multicolumn{2}{c}{Improvement(\%)} 
& 24.57 & 28.43 & 6.57 & - && 16.58  & 19.49 & 4.26 & - && 6.10 & 7.30 & 2.07 & -\\
\bottomrule
\end{tabular}
}
\end{threeparttable}
\label{tab:performance}
\end{table*}

\subsection{Performance Comparison}
\subsubsection{Cumulative and Annual Returns}
\begin{figure}[htbp]
% \vskip 0.2in
\begin{center}
\centerline{\includegraphics[width=0.75\columnwidth]{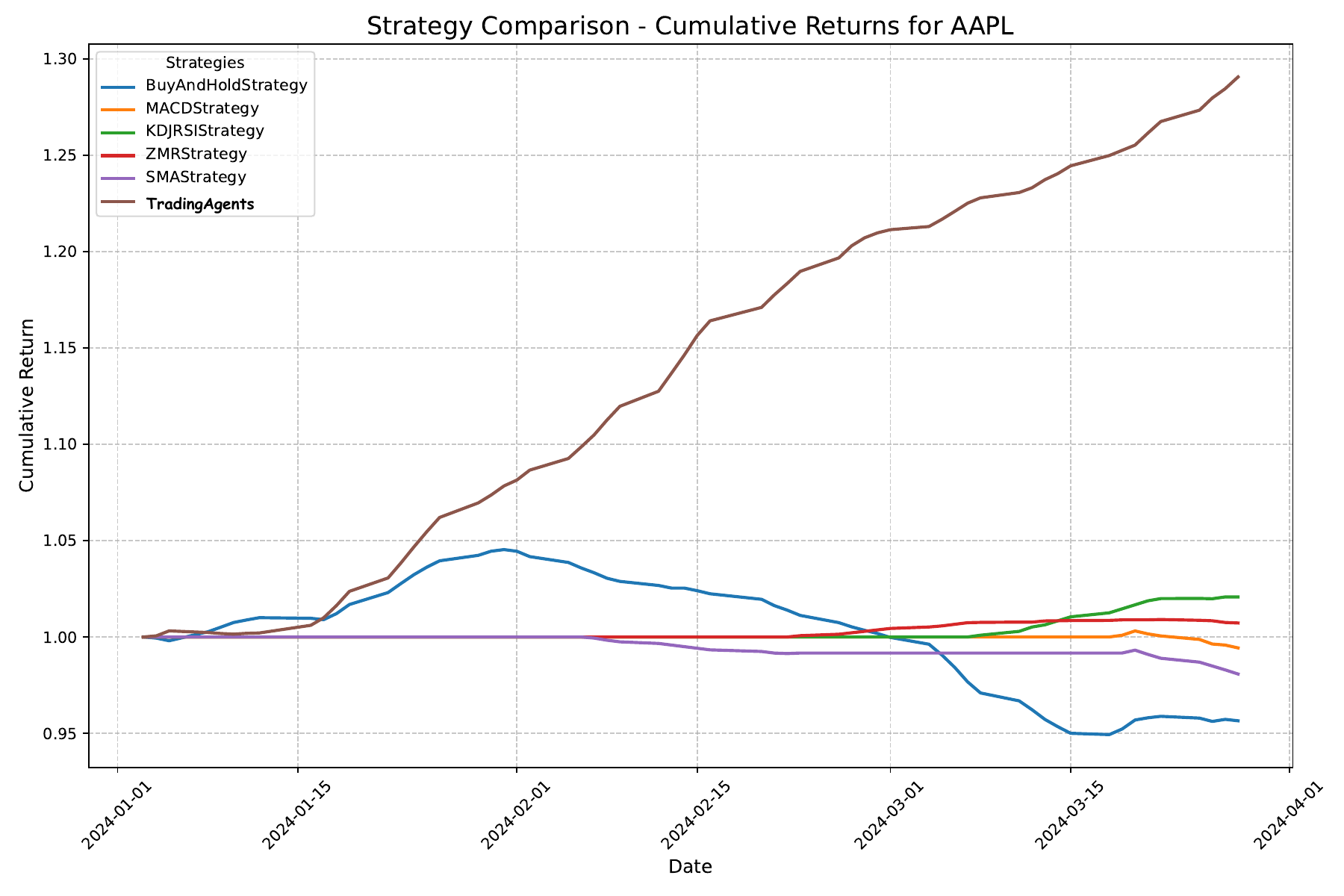}}
\caption{Cumulative Returns on \$AAPL using \textbf{\textcolor{brown}{\model}}. The figure shows the performance comparison of our model against baseline approaches for Apple Inc. stock analysis.}
\label{fig:aapl-cumulative-returns}
\end{center}
\vskip -0.2in
\end{figure}

\begin{figure}[htbp]
\vskip 0.2in
\begin{center}
\includegraphics[width=0.8\textwidth]{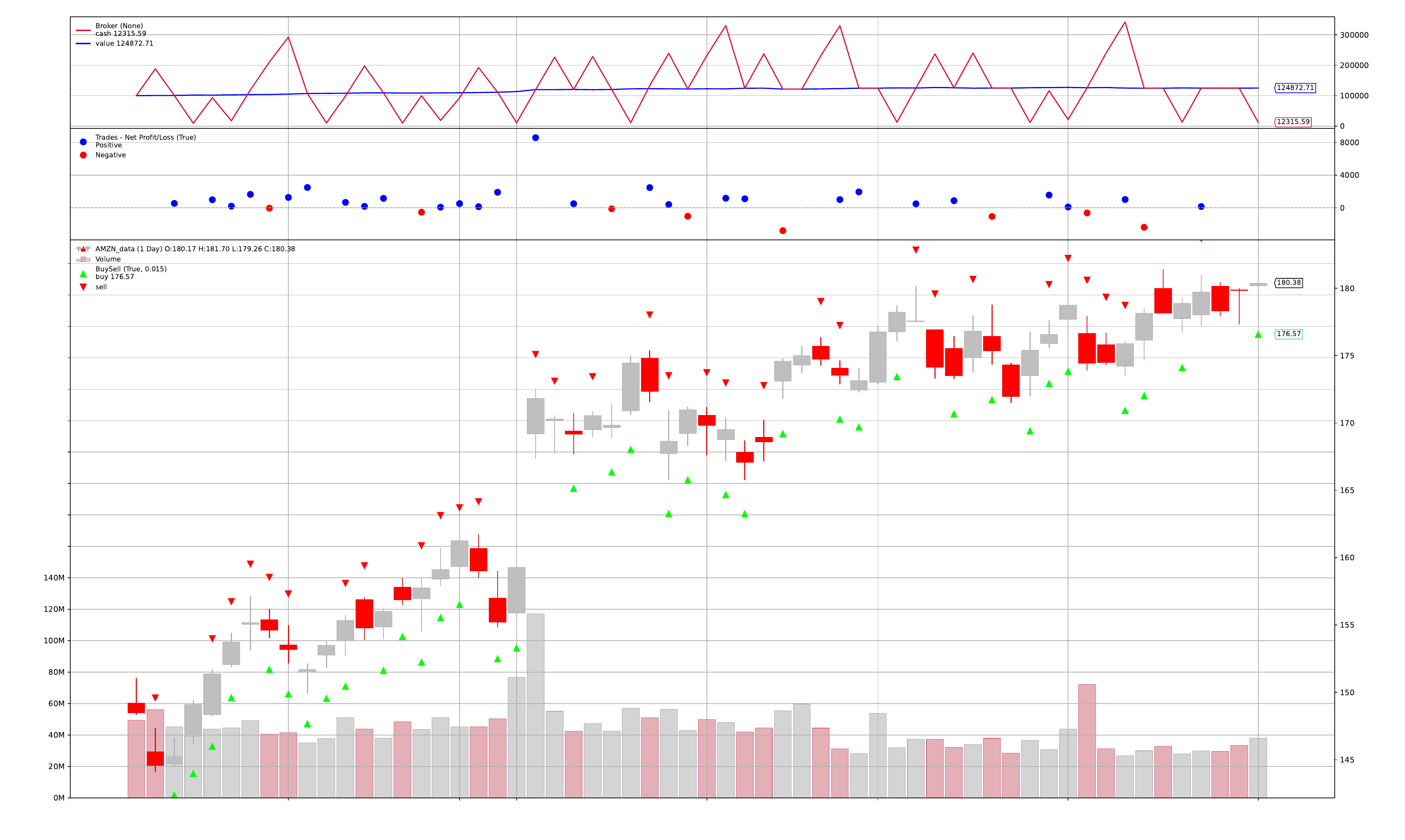}
\caption{\textbf{\textcolor{brown}{\model}}: Transaction History for AMZN.
Detailed transaction history with 
\textcolor{arr_green}{\texttt{Green}} \textcolor{brown}{\textbf{/}} \textcolor{arr_red}{\texttt{Red}} 
arrows for \textcolor{arr_green}{\texttt{Long}} \textcolor{brown}{\textbf{/}} \textcolor{arr_red}{\texttt{Short}} 
positions.}
\label{fig:amzn-performance-details}
\end{center}
\vskip -0.2in
\end{figure}

\begin{figure}[htbp]
\vskip 0.2in
\begin{center}
\includegraphics[width=0.8\textwidth]{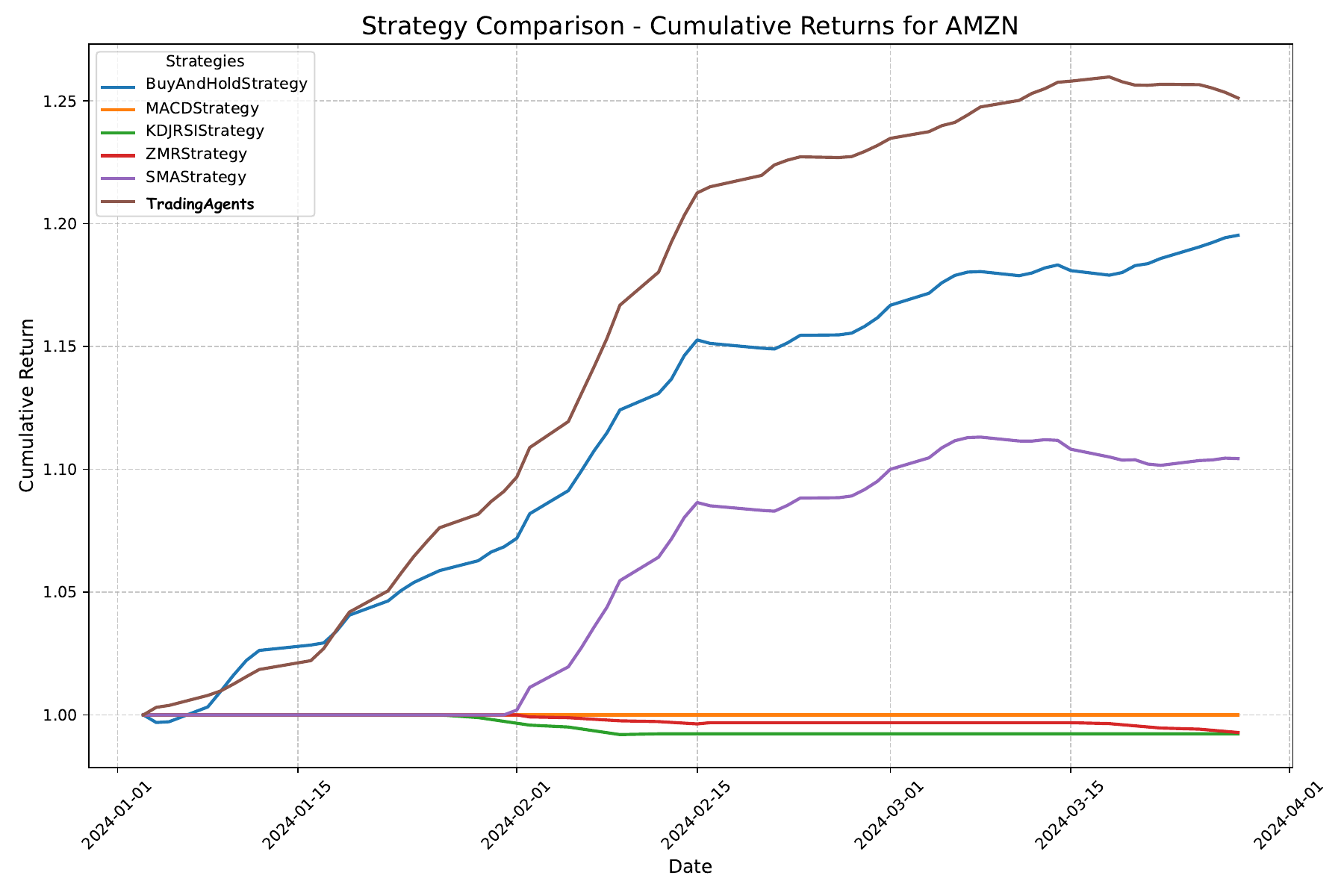}
\caption{\textbf{\textcolor{brown}{\model}}: Cumulative Returns for AMZN.
Performance comparison showing cumulative returns.}
\label{fig:amzn-performance-comparison}
\end{center}
\vskip -0.2in
\end{figure}

\begin{figure}[htbp]
\vskip 0.2in
\begin{center}
\includegraphics[width=0.8\textwidth]{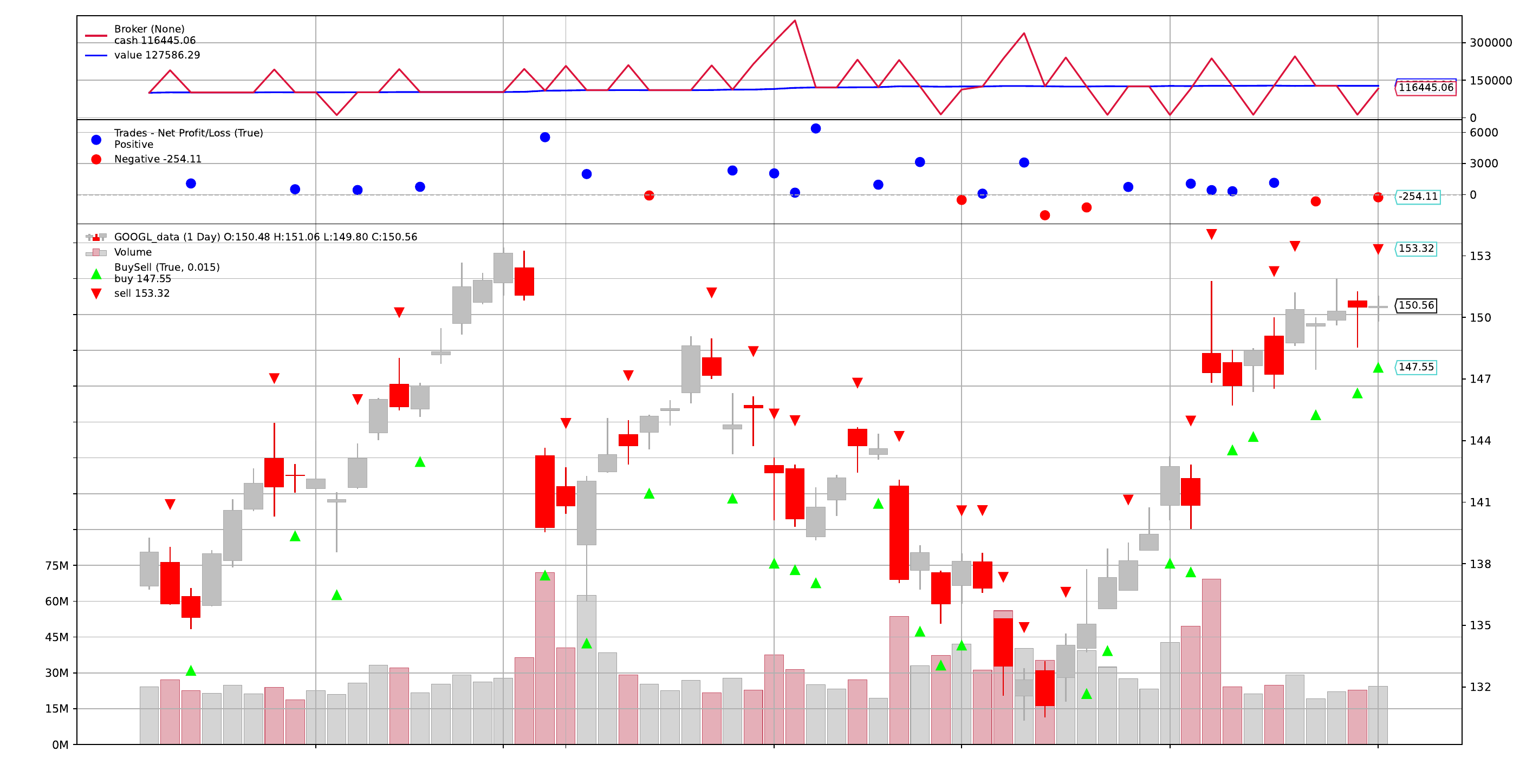}
\caption{\textbf{\textcolor{brown}{\model}}: Transaction History for GOOGL.
Detailed transaction history with 
\textcolor{arr_green}{\texttt{Green}} \textcolor{brown}{\textbf{/}} \textcolor{arr_red}{\texttt{Red}} 
arrows for \textcolor{arr_green}{\texttt{Long}} \textcolor{brown}{\textbf{/}} \textcolor{arr_red}{\texttt{Short}} 
positions.}
\label{fig:googl-performance-details}
\end{center}
\vskip -0.2in
\end{figure}

\begin{figure}[htbp]
\vskip 0.2in
\begin{center}
\includegraphics[width=0.8\textwidth]{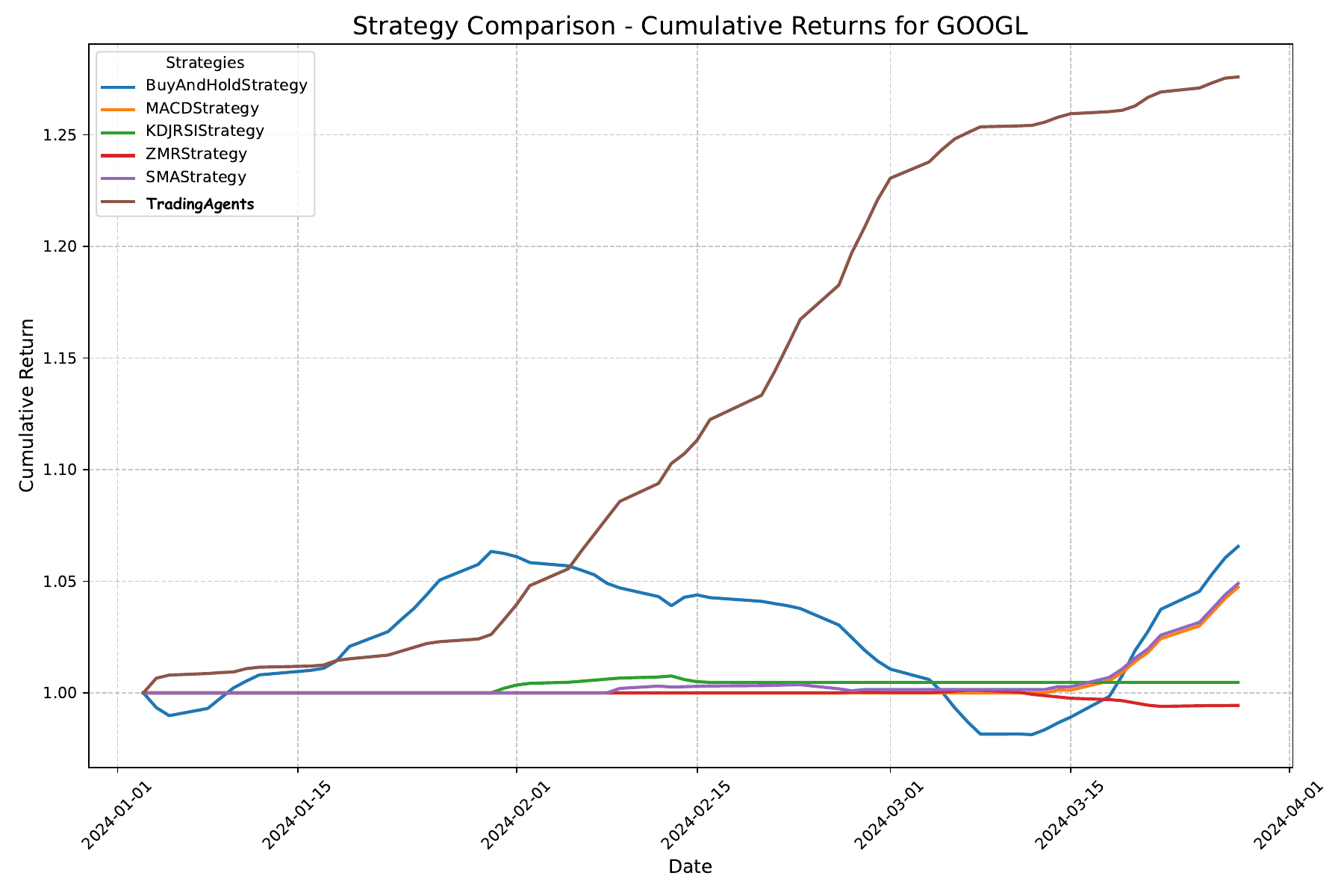}
\caption{\textbf{\textcolor{brown}{\model}}: Cumulative Returns for GOOGL.
Performance comparison showing cumulative returns.}
\label{fig:googl-performance-comparison}
\end{center}
\vskip -0.2in
\end{figure}

Table \ref{tab:performance} and Figures \ref{fig:aapl-transactions}, \ref{fig:aapl-cumulative-returns}, \ref{fig:amzn-performance-details}, \ref{fig:amzn-performance-comparison}, \ref{fig:googl-performance-details}, and \ref{fig:googl-performance-comparison} demonstrate that our method outperforms existing rule-based trading baselines, particularly in profitability, as measured by returns. \model achieves at least a 23.21\% cumulative return and a 24.90\% annual return on the three sampled stocks, surpassing the best-performing baselines by a margin of 6.1\%. Notably, on \$AAPL stock—a particularly challenging case due to market volatility during the testing period—traditional methods struggled, as their patterns failed to generalize to this situation. In contrast, \model excelled under these adverse conditions, achieving returns exceeding 26\% within months.

% ICML version

\vskip -1in

\subsubsection{Sharpe Ratio}

The Sharpe Ratio\footnote{We benchmarked \model over 3 months due to intensive LLM and tool use (11 LLM calls \& 20+ tool calls/prediction). The highest Sharpe Ratio exceeds our expected empirical range (\(SR\) above 2 -- very good, above 3 -- excellent). We exported \model's decision sequences and examined them to ensure calculation correctness. We believe the exceptionally high \(SR\) resulted from the phenomenon that there were few pullbacks in \model during that period. We report results as they are in our experiments faithfully. Future work will optimize LLM reasoning \& tool use to enable longer backtesting under limited budgets.}
 performance highlights \model's exceptional ability to deliver superior risk-adjusted returns, surpassing all baseline models. This result underscores \model's effectiveness in balancing returns and risk— a crucial factor for sustainable and predictable investment growth. \model outperforms market benchmarks such as Buy-and-Hold and rule-based strategies consistently, demonstrating its adaptability. Its capability to maximize returns while maintaining controlled risk exposure establishes a strong foundation for multi-agent and debate-based automated trading algorithms.

\subsubsection{Maximum Drawdown}
While rule-based baselines demonstrated superior performance in controlling risk, as reflected by their maximum drawdown scores, they fell short in capturing high returns. This trade-off between risk and reward underscores \model's strength as a balanced approach. Despite higher returns being typically associated with higher risks, \model maintained a relatively low maximum drawdown compared to many baselines. Its effective risk-control mechanisms, facilitated by the debates among risk-control agents, ensured that the maximum drawdown remained within a manageable limit, not exceeding 2. This demonstrates \model's capability to strike a robust balance between maximizing returns and managing risk effectively.

\subsubsection{Explainability}
A major drawback of current deep learning methods for trading is their dense, complex architectures, often rendering trading agents' decisions indecipherable. This challenge, rooted in AI explainability, is critical for trading agents operating in real-world financial markets, where incorrect decisions can cause severe losses. 

In contrast, an LLM-based agentic framework offers a transformative advantage: its decisions are communicated in natural language, enhancing interpretability. To illustrate, we provide \model's full trading log for a single day in the Appendix, showcasing its use of the ReAct-style prompting framework \citep{yao2023reactsynergizingreasoningacting}. Each decision includes detailed reasoning, tool usage, and thought processes, enabling traders to understand and debug the system. This transparency empowers traders to fine-tune the framework, accounting for decision factors, offering superior explainability over deep-learning trading algorithms.

\subsection{Discussion}
Our results demonstrate that integrating multiple specialized LLM agents and fostering agentic debate significantly enhances trading performance. This framework efficiently synthesizes diverse data sources and expert analyses, enabling trader agents to make well-informed decisions tailored to specific risk profiles. The inclusion of a reflective agent and a dedicated risk management team is pivotal in refining strategies and mitigating risks. As a result, the framework achieves exceptional return capture while maintaining strong risk management metrics, striking an optimal balance between maximizing rewards and minimizing risks. Additionally, the natural language-based operations of the multi-agent LLM framework ensure high explainability, giving \model a distinct advantage over traditional and deep learning methods in transparency and interoperability.

\section{Conclusion}

We introduced \model, a multi-agents LLM financial trading framework that realistically simulates a trading firm environment with multiple specialized agents engaging in agentic debates and conversations. Leveraging the advanced capabilities of LLMs to process and analyze diverse financial data sources, the framework enables more informed trading decisions while utilizing multi-agent interactions to enhance performance through comprehensive reasoning and debate before acting. By integrating agents with distinct roles and risk profiles, along with a reflective agent and a dedicated risk team, \model significantly improves trading outcomes and risk management compared to baseline models. Additionally, the collaborative nature of these agents ensures adaptability to varying market conditions. Experiments demonstrate that \model outperforms traditional trading strategies and baselines in cumulative return, Sharpe ratio, and other critical financial metrics. Future work will focus on deploying the framework in a live trading environment, expanding agent roles, and incorporating real-time data feeds to enhance performance further.

\clearpage
\newpage

\bibliographystyle{abbrvnat}
\bibliography{references}

% \section*{Acknowledgments} 
% Acknowledgments goes here.

\clearpage
\beginsupplement

\definecolor{lightblue}{rgb}{0.85, 0.92, 1.0}   % soft light blue, for background or highlights
\definecolor{lightgray}{rgb}{0.95, 0.95, 0.95}   % very light gray, for subtle contrasts
\definecolor{lightgreen}{rgb}{0.80, 0.92, 0.80}  % soft light green, for positive highlights
\definecolor{lightyellow}{rgb}{1.0, 1.0, 0.85}   % pale light yellow, for attention-catching highlights
\definecolor{lightpink}{rgb}{1.0, 0.85, 0.90}    % soft pink, for gentle emphasis
\definecolor{lightpurple}{rgb}{0.75, 0.65, 1.0}  % muted light purple, for distinguishing sections
\definecolor{lightorange}{rgb}{1.0, 0.90, 0.80}  % light orange, for calls to action or warnings
\definecolor{lightred}{rgb}{1.0, 0.80, 0.80}     % light red, for critical notes or errors
\definecolor{brightblue}{rgb}{0.45, 0.65, 1.0}   % medium bright blue, for links or links to references
\definecolor{lightlilac}{rgb}{0.85, 0.75, 1.0}  % soft lilac, for decorative or neutral accents

\lstdefinestyle{agent1}{
    backgroundcolor=\color{lightblue}, 
    basicstyle=\ttfamily\footnotesize,
    frame=single,
    rulecolor=\color{black},
    xleftmargin=0.5cm,
    framexleftmargin=0.5cm,
    numbers=left,
    numberstyle=\tiny\color{black},
    stepnumber=1,
    numbersep=5pt,
    lineskip=1pt,
    breaklines=true,
    breakatwhitespace=false
}

\lstdefinestyle{agent2}{
    backgroundcolor=\color{lightgray}, 
    basicstyle=\ttfamily\footnotesize,
    frame=single,
    rulecolor=\color{black},
    xleftmargin=0.5cm,
    framexleftmargin=0.5cm,
    numbers=left,
    numberstyle=\tiny\color{black},
    stepnumber=1,
    numbersep=5pt,
    lineskip=1pt,
    breaklines=true,
    breakatwhitespace=false
}

\lstdefinestyle{agent3}{
    backgroundcolor=\color{lightgreen}, 
    basicstyle=\ttfamily\footnotesize,
    frame=single,
    rulecolor=\color{black},
    xleftmargin=0.5cm,
    framexleftmargin=0.5cm,
    numbers=left,
    numberstyle=\tiny\color{black},
    stepnumber=1,
    numbersep=5pt,
    lineskip=1pt,
    breaklines=true,
    breakatwhitespace=false
}

\lstdefinestyle{agent4}{
    backgroundcolor=\color{lightyellow}, 
    basicstyle=\ttfamily\footnotesize,
    frame=single,
    rulecolor=\color{black},
    xleftmargin=0.5cm,
    framexleftmargin=0.5cm,
    numbers=left,
    numberstyle=\tiny\color{black},
    stepnumber=1,
    numbersep=5pt,
    lineskip=1pt,
    breaklines=true,
    breakatwhitespace=false
}

\lstdefinestyle{agent5}{
    backgroundcolor=\color{lightpink}, 
    basicstyle=\ttfamily\footnotesize,
    frame=single,
    rulecolor=\color{black},
    xleftmargin=0.5cm,
    framexleftmargin=0.5cm,
    numbers=left,
    numberstyle=\tiny\color{black},
    stepnumber=1,
    numbersep=5pt,
    lineskip=1pt,
    breaklines=true,
    breakatwhitespace=false
}

\lstdefinestyle{agent6}{
    backgroundcolor=\color{lightpurple}, 
    basicstyle=\ttfamily\footnotesize,
    frame=single,
    rulecolor=\color{black},
    xleftmargin=0.5cm,
    framexleftmargin=0.5cm,
    numbers=left,
    numberstyle=\tiny\color{black},
    stepnumber=1,
    numbersep=5pt,
    lineskip=1pt,
    breaklines=true,
    breakatwhitespace=false
}

\lstdefinestyle{agent7}{
    backgroundcolor=\color{lightorange}, 
    basicstyle=\ttfamily\footnotesize,
    frame=single,
    rulecolor=\color{black},
    xleftmargin=0.5cm,
    framexleftmargin=0.5cm,
    numbers=left,
    numberstyle=\tiny\color{black},
    stepnumber=1,
    numbersep=5pt,
    lineskip=1pt,
    breaklines=true,
    breakatwhitespace=false
}

\lstdefinestyle{agent8}{
    backgroundcolor=\color{lightred}, 
    basicstyle=\ttfamily\footnotesize,
    frame=single,
    rulecolor=\color{black},
    xleftmargin=0.5cm,
    framexleftmargin=0.5cm,
    numbers=left,
    numberstyle=\tiny\color{black},
    stepnumber=1,
    numbersep=5pt,
    lineskip=1pt,
    breaklines=true,
    breakatwhitespace=false
}

\lstdefinestyle{agent9}{
    backgroundcolor=\color{brightblue}, 
    basicstyle=\ttfamily\footnotesize,
    frame=single,
    rulecolor=\color{black},
    xleftmargin=0.5cm,
    framexleftmargin=0.5cm,
    numbers=left,
    numberstyle=\tiny\color{black},
    stepnumber=1,
    numbersep=5pt,
    lineskip=1pt,
    breaklines=true,
    breakatwhitespace=false
}

\lstdefinestyle{agent10}{
    backgroundcolor=\color{lightlilac}, 
    basicstyle=\ttfamily\footnotesize,
    frame=single,
    rulecolor=\color{black},
    xleftmargin=0.5cm,
    framexleftmargin=0.5cm,
    numbers=left,
    numberstyle=\tiny\color{black},
    stepnumber=1,
    numbersep=5pt,
    lineskip=1pt,
    breaklines=true,
    breakatwhitespace=false
}

% \newpage
\onecolumn

\section{Supplementary Materials for \model}

\subsection{Baseline Models}\label{appendix:baselines}

We compare our \model framework against several baselines:
\begin{itemize}

    \item \textbf{Buy and Hold}: Investing equal amounts in all selected stocks and holding them throughout the simulation period.

    \item \textbf{MACD (Moving Average Convergence Divergence)}: A trend-following momentum strategy that generates buy and sell signals based on the crossover points between the MACD line and signal line.

    \item \textbf{KDJ and RSI (Relative Strength Index)}: A momentum strategy combining KDJ (stochastic oscillator) and RSI (relative strength index) indicators to identify overbought and oversold conditions for trading signals.

    \item \textbf{ZMR (Zero Mean Reversion)}: A mean reversion trading strategy that generates signals based on price deviations from and subsequent reversions to a zero reference line.

    \item \textbf{SMA (Simple Moving Average)}: A trend-following strategy that generates trading signals based on crossovers between short-term and long-term moving averages.
\end{itemize}

\subsection{Evaluation Metrics}\label{appendix:eval_metrics}

To thoroughly evaluate the performance of our \model framework, we use widely recognized metrics to assess the risk management, profitability, and safety of the \model strategy in comparison to baseline approaches. Here we describe these metrics:

\subsubsection{Cumulative Return (CR)}

The cumulative return measures the total return generated over the simulation period. It is calculated as:

\begin{equation}
\text{CR} = \left( \frac{V_{\text{end}} - V_{\text{start}}}{V_{\text{start}}} \right) \times 100\%
\end{equation}

where $V_{\text{end}}$ is the portfolio value at the end of the simulation, and $V_{\text{start}}$ is the initial portfolio value.

\subsubsection{Annualized Return (AR)}

The annualized return normalizes the cumulative return over the number of years:

\begin{equation}
\text{AR} = \left( \left( \frac{V_{\text{end}}}{V_{\text{start}}} \right)^{\frac{1}{N}} - 1 \right) \times 100\%
\end{equation}

where $N$ is the number of years in the simulation.

\subsubsection{Sharpe Ratio (SR)}

The Sharpe ratio measures risk-adjusted return by comparing a portfolio's excess return over the risk-free rate to its volatility:

\begin{equation}
\text{SR} = \frac{\bar{R} - R_f}{\sigma}
\end{equation}

where $\bar{R}$ is the average portfolio return, $R_f$ is the risk-free rate (e.g., yield of 3-month Treasury bills), and $\sigma$ is the standard deviation of the portfolio returns.

\subsubsection{Maximum Drawdown (MDD)}

Maximum drawdown measures the largest peak-to-trough decline in the portfolio value:

\begin{equation}
\text{MDD} = \max_{t \in [0, T]} \left( \frac{\text{Peak}_t - \text{Trough}_t}{\text{Peak}_t} \right) \times 100\%
\end{equation}

\subsection{Cumulative Returns (CR) and Transaction History for AMZN and GOOGL}

We present additional figures for \verb|$AMZN| and \verb|$GOOGL| stocks to complement the AAPL data discussed in the main body of this paper. These supplementary visuals provide a broader perspective on the performance of our trading framework across multiple stocks, highlighting the consistency and robustness of \model's results.

By including detailed analyses for AMZN and GOOGL, we aim to demonstrate the versatility of our approach in diverse market environments, thereby reinforcing the overall effectiveness and generalizability of our methodology.

% \newpage

\subsection{\model Workflow: Role Specification and Cooperation} \label{app:trade_log}

We offer a comprehensive overview of the various agent roles that collaborate within the \model. These roles include the Analyst Team, Researcher Team, Trader, Risk Management Team, and Fund Manager, each dedicated to different aspects of the trading process for Apple Inc. (Ticker: AAPL) as of November 19, 2024, a recent date chosen for demonstration purposes.

By detailing the responsibilities and interactions of each agent, we demonstrate how \model ensures thorough analysis, strategic trading decisions, and effective risk management. The collaborative structure not only enhances the system's overall performance but also reinforces the transparency and explainability of \model, aligning with the objectives discussed in the main text. The following sections provide specific role specifications and illustrate how these agents work together to achieve robust and interpretable trading outcomes.

\definecolor{lightblue}{rgb}{0.90, 0.95, 1.0}
\definecolor{lightbeige}{rgb}{0.98, 0.96, 0.92}
\definecolor{lightgreen}{rgb}{0.88, 0.95, 0.88}
\definecolor{lightyellow}{rgb}{1.0, 1.0, 0.90}
\definecolor{lightpink}{rgb}{1.0, 0.90, 0.94}
\definecolor{lightpurple}{rgb}{0.85, 0.80, 1.0}
\definecolor{lightorange}{rgb}{1.0, 0.94, 0.86}
\definecolor{lightred}{rgb}{1.0, 0.86, 0.86}
\definecolor{lightmint}{rgb}{0.92, 1.0, 0.96}
\definecolor{lightlilac}{rgb}{0.90, 0.85, 1.0}

% Base lstlisting style (no frame, for use inside tcolorbox)
\lstdefinestyle{agentbase}{
    basicstyle=\ttfamily\footnotesize,
    frame=none,
    xleftmargin=0cm,
    numbers=none,
    breaklines=true,
    breakatwhitespace=false,
}

% Market Analyst Team - Light Mint
\begin{tcolorbox}[
    breakable,
    enhanced,
    colback=lightmint!70!white,
    colframe=teal!20!gray,
    coltitle=black,
    boxrule=0.5pt,
    arc=0.5mm,
    top=2mm, bottom=2mm, left=2mm, right=2mm,
    fonttitle=\usefont{T1}{ppl}{b}{n},
    title=\model: Market Analyst Team -- Apple (AAPL),
]
\usefont{T1}{ppl}{m}{n}
\footnotesize
\begin{lstlisting}[style=agentbase]
[Market Analyst]:
================================== Report ==================================
Based on the analysis of AAPL's recent market data and selected technical indicators, here is a detailed and nuanced report of the trends observed:

### Momentum Indicators
- **RSI (Relative Strength Index)**: The RSI values indicate significant fluctuations, with periods of overbought conditions, particularly on November 5 and November 7, where RSI values reached 100 and 89.85, respectively. This suggests strong upward price movements, potentially leading to price corrections. The RSI stabilizes towards the end of the period, indicating moderate buying momentum.

### Trend Indicators
- **ADX (Average Directional Index)**: Initially, the ADX values are extremely high, indicating a strong trend, but they decline significantly, suggesting a weakening trend strength. This pattern implies a transition from a strong directional movement to a potential consolidation phase.

- **Supertrend**: The supertrend indicator remains constant, suggesting a stable market trend without significant volatility. This stability might indicate a consolidation phase, with the potential for future breakouts.

### Volatility Indicators
- **Bollinger Bands**: The upward trend in Bollinger Bands suggests a bullish market phase with low volatility. The consistent rise in values indicates a positive market bias, with prices likely to continue rising if the trend persists.

- **ATR (Average True Range)**: The ATR values show an increase in market volatility, peaking on November 7. This suggests heightened market activity, with sustained volatility towards the end of the period.

### Volume Indicators
- **VR (Volume Variation Index)**: The VR values highlight significant fluctuations in trading volume, with a notable spike on November 5. This indicates heightened market activity, possibly due to external factors influencing trader behavior.

### Price Action and Support/Resistance
- **CCI (Commodity Channel Index)**: The CCI values indicate volatility, with several instances of overbought conditions. Peaks on November 7, November 14, and November 18 suggest strong upward price movements, which could be followed by corrections.

### Moving Averages
- **MACD (Moving Average Convergence Divergence)**: The MACD values indicate a general upward trend, with some fluctuations and minor pullbacks. The increasing MACD values suggest bullish market sentiment, with the short-term moving average consistently above the long-term average.

### Summary and Insights
The analysis of AAPL's recent market data reveals a period of volatility with alternating phases of strong buying momentum and stabilization. The initial overbought conditions and high ADX values suggest a strong trend, which weakens over time, indicating potential consolidation. The consistent upward trend in Bollinger Bands and MACD values points to a bullish market sentiment, while the fluctuations in ATR and VR values highlight ongoing market volatility.

Traders should be cautious of potential price corrections following the overbought conditions and consider the possibility of continued volatility. The stable supertrend value suggests a consolidation phase, with potential for future breakouts. Overall, traders should remain vigilant, using these insights to adjust their strategies accordingly, while being prepared for both opportunities and risks presented by the current market conditions.
================================ Tool Calls ================================
Tool Calls:
  get_YFin_data (call_TuDo4m3U0lkgtV61Qpq3UIiT)
 Call ID: call_TuDo4m3U0lkgtV61Qpq3UIiT
  Args:
    symbol: AAPL
    curr_date: 2024-11-19

Based on the recent market data for Apple (AAPL) from November 4, 2024, to November 18, 2024, we can observe several key trends and patterns:

1. **Price Movement**: AAPL has shown a general upward trend over the period, with the stock price increasing from an opening of
$220.75 on November 4 to a closing of $228.02 on November 18. The stock experienced fluctuations, with notable highs on November
7 and November 14, reaching $227.63 and $228.87, respectively.

2. **Volume Trends**: Trading volume varied significantly, with a peak on November 6 at 54,561,100 shares. This suggests heightened trading activity, possibly due to market events or investor sentiment changes.

3. **Dividend Payment**: A dividend of $0.25 per share was paid on November 8, which might have influenced investor behavior, potentially contributing to the price movements observed around this date.

4. **Volatility**: The stock showed some volatility, with daily price ranges (highs and lows) indicating active trading and potential market reactions to news or earnings reports.

To provide a comprehensive analysis, I will select a set of technical indicators that can offer diverse insights into AAPL's market behavior. These indicators will cover momentum, trend, volatility, volume, and price action aspects:

1. **RSI (Relative Strength Index)**: To assess momentum and identify overbought or oversold conditions.
2. **ADX (Average Directional Index)**: To evaluate the strength of the current trend.
3. **Bollinger Bands**: To analyze volatility and potential price breakouts.
4. **MACD (Moving Average Convergence Divergence)**: To identify trend changes and momentum.
5. **VWMA (Volume Weighted Moving Average)**: To understand price movements in relation to volume.
6. **ATR (Average True Range)**: To measure market volatility.
7. **Supertrend**: To identify trend direction and potential reversals.
8. **CCI (Commodity Channel Index)**: To identify cyclical trends and potential reversals.

I will now retrieve detailed reports for these indicators to provide a nuanced analysis of AAPL's market conditions.
Tool Calls:
  get_stockstats_indicators_report (call_rARQhoVdv6VyzI3a3R9hyhYh)
 Call ID: call_rARQhoVdv6VyzI3a3R9hyhYh
  Args:
    symbol: AAPL
    indicator: rsi
    curr_date: 2024-11-19
  get_stockstats_indicators_report (call_ODamPSS2si621zrEPb4mIitD)
 Call ID: call_ODamPSS2si621zrEPb4mIitD
  Args:
    symbol: AAPL
    indicator: adx
    curr_date: 2024-11-19
  get_stockstats_indicators_report (call_KWgyXj3GgdviC01u5KeFZkbQ)
 Call ID: call_KWgyXj3GgdviC01u5KeFZkbQ
  Args:
    symbol: AAPL
    indicator: boll
    curr_date: 2024-11-19
  get_stockstats_indicators_report (call_hshUKMGu1yvZ4m8YbJE9SaSg)
 Call ID: call_hshUKMGu1yvZ4m8YbJE9SaSg
  Args:
    symbol: AAPL
    indicator: macd
    curr_date: 2024-11-19
  get_stockstats_indicators_report (call_Iq9kq9jA2D5E7Png9lPt9bK5)
 Call ID: call_Iq9kq9jA2D5E7Png9lPt9bK5
  Args:
    symbol: AAPL
    indicator: vwma
    curr_date: 2024-11-19
  get_stockstats_indicators_report (call_SRSi644IhPjxaMdE7U8xn4a3)
 Call ID: call_SRSi644IhPjxaMdE7U8xn4a3
  Args:
    symbol: AAPL
    indicator: atr
    curr_date: 2024-11-19
  get_stockstats_indicators_report (call_Ow6VLKfP52JSFtphkJocda9f)
 Call ID: call_Ow6VLKfP52JSFtphkJocda9f
  Args:
    symbol: AAPL
    indicator: supertrend
    curr_date: 2024-11-19
  get_stockstats_indicators_report (call_FJYy18HBaxQlC8YghPphXpsB)
 Call ID: call_FJYy18HBaxQlC8YghPphXpsB
  Args:
    symbol: AAPL
    indicator: cci
    curr_date: 2024-11-19

The data provided spans from November 4, 2024, to November 18, 2024, and includes Commodity Channel Index (CCI) values, which measure the deviation of the price from its average. The CCI values indicate the following trends and insights:

1. **Initial Missing Data**: On November 4, 2024, the CCI value is missing. This absence of data could be due to a lack of trading activity or data collection issues, which should be noted but does not provide insight into market conditions.

2. **Positive Trend**: From November 5 to November 7, there is a noticeable upward trend in CCI values, starting at 66.67 and peaking at 133.33. This suggests a strong deviation from the average price, indicating potential overbought conditions. Traders might interpret this as a signal that the market could be due for a correction or pullback.

3. **Fluctuations and Correction**: On November 8, the CCI decreases to 97.80, indicating a slight correction from the previous peak. This is followed by a significant drop to -9.75 on November 11, suggesting a shift towards oversold conditions. Such a rapid change could indicate increased volatility or a reaction to external market factors.

4. **Recovery and Stabilization**: From November 12 to November 14, the CCI values recover, moving from 10.83 to 109.26. This recovery suggests a return to positive momentum, potentially indicating renewed buying interest or stabilization in the market.

5. **Moderate Fluctuations**: On November 15, the CCI drops to 30.79, followed by an increase to 104.68 on November 18. These fluctuations suggest ongoing market adjustments, with the CCI remaining in positive territory, indicating that the market is still experiencing deviations above the average price.

Overall, the data reflects a period of volatility with significant swings in market sentiment. The initial rise and subsequent fluctuations in CCI values suggest that traders should be cautious of potential overbought or oversold conditions. Monitoring these trends can help in making informed decisions about potential entry or exit points in the market.
\end{lstlisting}
\end{tcolorbox}

% News Analyst Team - Light Blue
\begin{tcolorbox}[
    breakable,
    enhanced,
    colback=lightblue!50!white,
    colframe=brightblue!30!gray,
    coltitle=black,
    boxrule=0.5pt,
    arc=0.5mm,
    top=2mm, bottom=2mm, left=2mm, right=2mm,
    fonttitle=\usefont{T1}{ppl}{b}{n},
    title=\model: News Analyst Team -- Apple (AAPL),
]
\usefont{T1}{ppl}{m}{n}
\footnotesize
\begin{lstlisting}[style=agentbase]
[News Analyst]:
================================== Report ==================================
Based on the recent news from EODHD and Finnhub, here is a comprehensive report on the current state of the world relevant for trading and macroeconomics:

### Global Economic and Political Environment
1. **US-China Relations and Trade**: There is ongoing tension between the US and China, particularly affecting companies like Apple that have significant operations in China. Jim Cramer highlighted the challenges of doing business in China amidst political tensions, which could impact Apple's operations and stock performance.

2. **US Economic Policy**: The return of Donald Trump to the presidency is creating mixed reactions in the business community. Some tech executives are optimistic about potential deregulation and increased innovation, which could lead to more spending and dealmaking.

3. **Inflation and Federal Reserve Policy**: The October CPI inflation report showed a rise to 2.6%, indicating that the Federal Reserve's management of inflation is on track. However, this has increased the likelihood of a pause in rate cuts, affecting market expectations and investor sentiment.

### Sector-Specific Insights
1. **Technology and AI**:
- **Apple**: Apple is expanding into the smart home market with a new AI-powered device, competing with Amazon and Google. Despite strong earnings, there are concerns about overvaluation and challenges in China.
- **Google (Alphabet)**: Google is facing pressure from AI competition in the search ad business and ongoing DOJ lawsuits. Analysts have mixed views on its stock, with some not considering it a favored stock currently.
- **Amazon**: Amazon is making significant moves into full autonomy and expanding its AI services through AWS. However, its stock recently declined due to broader market corrections and Federal Reserve comments on interest rates.

2. **Semiconductors**:
- **Taiwan Semiconductor Manufacturing Company (TSMC)**: The US has finalized a $6.6 billion subsidy for TSMC's semiconductor production in Arizona, highlighting the strategic importance of semiconductor manufacturing. Despite high demand driven by AI, TSMC is considered overvalued by some analysts due to geopolitical risks and its role in US-China economic competition.

3. **Investment Trends**:
- **AI and Tech Stocks**: There is a continued focus on AI-centric tech giants like Google, Meta, TSMC, Adobe, and Broadcom for potential double-digit returns. The rapid adoption of generative AI is likened to the growth of Facebook, indicating significant investment opportunities in AI technologies.

### Market Sentiment and Investor Behavior
1. **Portfolio Management**: Major investment firms like Fisher Asset Management and GMO Capital have shown increased portfolio values, with significant holdings in tech giants such as Apple, Microsoft, Nvidia, Amazon, and Alphabet.

2. **Investor Interest**: There is heightened investor interest in Apple, as indicated by increased searches and discussions about its stock prospects. This reflects broader market attention on tech stocks amidst evolving economic conditions.

### Conclusion
The current global economic landscape is characterized by geopolitical tensions, particularly between the US and China, impacting multinational companies. The technology sector, especially AI and semiconductor industries, remains a focal point for investors, with significant government support and market interest. Inflation management by the Federal Reserve continues to influence market dynamics, affecting investor expectations and stock performance. Traders should consider these factors when making investment decisions, focusing on sectors with strong growth potential and navigating geopolitical risks.
================================ Tool Calls ================================
Tool Calls:
  get_EODHD_news (call_RCXogBfhGwyi0IuWCWhpNpQJ)
 Call ID: call_RCXogBfhGwyi0IuWCWhpNpQJ
  Args:
    start_date: 2024-11-12
    end_date: 2024-11-19
  get_finnhub_news (call_zXlMVeXHxfA2X4bDTXmIfeij)
 Call ID: call_zXlMVeXHxfA2X4bDTXmIfeij
  Args:
    ticker: AAPL
    start_date: 2024-11-12
    end_date: 2024-11-19
  get_finnhub_news (call_jSKV1zMfHVqrR4GYAeEzjsoy)
 Call ID: call_jSKV1zMfHVqrR4GYAeEzjsoy
  Args:
    ticker: TSM
    start_date: 2024-11-12
    end_date: 2024-11-19
  get_finnhub_news (call_7rrmR2Qo5K8O2WbxQc8zDpxB)
 Call ID: call_7rrmR2Qo5K8O2WbxQc8zDpxB
  Args:
    ticker: GOOGL
    start_date: 2024-11-12
    end_date: 2024-11-19
  get_finnhub_news (call_XmPHBNBRJINGDZmBeUSnUgps)
 Call ID: call_XmPHBNBRJINGDZmBeUSnUgps
  Args:
    ticker: AMZN
    start_date: 2024-11-12
    end_date: 2024-11-19
\end{lstlisting}
\end{tcolorbox}

% Social Media Analysts Team - Light Beige
\begin{tcolorbox}[
    breakable,
    enhanced,
    colback=lightbeige!70!white,
    colframe=brown!20!gray,
    coltitle=black,
    boxrule=0.5pt,
    arc=0.5mm,
    top=2mm, bottom=2mm, left=2mm, right=2mm,
    fonttitle=\usefont{T1}{ppl}{b}{n},
    title=\model: Social Media Analysts Team -- Apple (AAPL),
]
\usefont{T1}{ppl}{m}{n}
\footnotesize
\begin{lstlisting}[style=agentbase]
[Social Media Analyst]:
================================== Report ==================================
### Comprehensive Analysis of Apple Inc. (AAPL) Social Media and Sentiment Trends (November 12 - November 19, 2024)

#### Social Media Insights
Upon reviewing the Reddit data for Apple Inc. over the past week, it appears there were no significant posts or discussions captured in the dataset. This absence of data could suggest a lack of major events or announcements that typically drive social media engagement, or it might indicate a gap in data collection. For investors, this means relying more heavily on sentiment analysis and news reports for insights during this period.

#### Sentiment Analysis
The sentiment data for Apple Inc. (AAPL) from November 4, 2024, to November 17, 2024, reveals a dynamic landscape of public perception:

1. **Positive Sentiment Peaks**:
- **November 15, 2024**: A significant positive sentiment was recorded with a normalized score of 0.5445, indicating favorable news or developments.
- **November 11, 2024**: Another strong positive sentiment day with a score of 0.426, suggesting positive market reactions or announcements.

2. **Moderate Positive Sentiment**:
- Several days, including November 14, 13, 10, and 9, showed moderately positive sentiment, reflecting a generally favorable outlook towards AAPL.

3. **Negative Sentiment**:
- **November 12, 2024**: A negative sentiment score of -0.201 was observed, possibly due to unfavorable news or market conditions.
- **November 4, 2024**: The period started with a negative sentiment score of -0.141, indicating initial concerns or negative news.

4. **Neutral Sentiment**:
- **November 17, 2024**: The sentiment was neutral, suggesting a lack of significant positive or negative news impacting perceptions.

5. **Engagement Levels**:
- The highest engagement was noted on November 6, 2024, with a sentiment score of 0.0756, indicating increased interest or discussions, albeit with a slightly positive sentiment.

#### Implications for Traders and Investors
- **Positive Sentiment Days**: The spikes in positive sentiment, particularly on November 15 and 11, could be indicative of favorable developments or market conditions that investors might want to capitalize on.
- **Negative Sentiment Days**: The negative sentiment on November 12 and 4 suggests caution, as these days might have been influenced by adverse news or market reactions.
- **Overall Trend**: The sentiment trend over the period is predominantly positive, suggesting a generally favorable outlook for AAPL. However, the occasional dips into negative sentiment highlight the importance of staying informed about potential risks or negative news.

In conclusion, while the lack of Reddit data limits insights from social media discussions, the sentiment analysis provides a valuable perspective on market perceptions of Apple Inc. during this period. Investors should consider these sentiment trends alongside other market data and news to make informed decisions.
================================ Tool Calls ================================
Tool Calls:
  get_reddit_stock_info (call_MfJT2mtj22GO32q7Xu5z2htS)
 Call ID: call_MfJT2mtj22GO32q7Xu5z2htS
  Args:
    query: Apple Inc
    start_date: 2024-11-12
    end_date: 2024-11-19
  get_EODHD_sentiment (call_So5Csrkq4CJKpi9jiY4ibbB2)
 Call ID: call_So5Csrkq4CJKpi9jiY4ibbB2
  Args:
    symbol: AAPL
    curr_date: 2024-11-19
EODHD client initialized
Reddit client initialized
Searching in subreddit: wallstreetbets
Searching in subreddit: stocks
Searching in subreddit: investing
Searching in subreddit: SecurityAnalysis
Searching in subreddit: Finance
Searching in subreddit: Economics

The market sentiment data for AAPL over the period from November 4, 2024, to November 17, 2024, provides insights into the prevailing attitudes and perceptions surrounding the stock. Below is a detailed analysis of the sentiment trends observed during this period:

1. **November 17, 2024**: The sentiment count was 1, with a normalized sentiment score of 0. This indicates a neutral sentiment on this day, suggesting no significant positive or negative sentiment was detected.

2. **November 16, 2024**: The sentiment count was 1, with a normalized sentiment score of -0.06. This reflects a slightly negative sentiment, although the impact appears minimal given the low count and score.

3. **November 15, 2024**: The sentiment count increased to 2, with a normalized sentiment score of 0.5445. This marks a significant positive sentiment, indicating a favorable perception of AAPL on this day.

4. **November 14, 2024**: The sentiment count was 2, with a normalized sentiment score of 0.17. This suggests a moderately positive sentiment, showing a generally favorable outlook.

5. **November 13, 2024**: The sentiment count was 2, with a normalized sentiment score of 0.3245. This indicates a positive sentiment, with a noticeable improvement in perception compared to the previous day.

6. **November 12, 2024**: The sentiment count was 2, with a normalized sentiment score of -0.201. This reflects a negative sentiment, suggesting some unfavorable views or news impacting AAPL.

7. **November 11, 2024**: The sentiment count was 3, with a normalized sentiment score of 0.426. This indicates a strong positive sentiment, with increased engagement and favorable perception.

8. **November 10, 2024**: The sentiment count was 3, with a normalized sentiment score of 0.3157. This suggests a positive sentiment, maintaining a favorable outlook similar to the previous day.

9. **November 09, 2024**: The sentiment count was 3, with a normalized sentiment score of 0.221. This indicates a positive sentiment, though slightly less pronounced than the previous days.

10. **November 08, 2024**: The sentiment count was 1, with a normalized sentiment score of 0.422. This reflects a strong positive sentiment, despite the lower engagement count.

11. **November 07, 2024**: The sentiment count was 2, with a normalized sentiment score of 0.144. This suggests a mildly positive sentiment, indicating a generally favorable view.

12. **November 06, 2024**: The sentiment count was 5, with a normalized sentiment score of 0.0756. This shows a slightly positive sentiment with the highest engagement count, suggesting increased interest or discussion around AAPL.

13. **November 05, 2024**: The sentiment count was 4, with a normalized sentiment score of 0.304. This indicates a positive sentiment, with a relatively high level of engagement.

14. **November 04, 2024**: The sentiment count was 3, with a normalized sentiment score of -0.141. This reflects a negative sentiment, suggesting some unfavorable sentiment or news at the start of the period.

**Summary**: Over the analyzed period, AAPL experienced fluctuations in market sentiment, with notable positive sentiment spikes on November 15 and November 11. The sentiment was predominantly positive, with occasional dips into negative territory, particularly on November 12 and November 04. The engagement levels varied, with the highest count observed on November 06, indicating heightened interest or discussion on that day. Overall, the sentiment trend suggests a generally favorable outlook for AAPL during this period, with some days of negative sentiment that may warrant further investigation.
\end{lstlisting}
\end{tcolorbox}

% Fundamentals Analyst Team - Light Green
\begin{tcolorbox}[
    breakable,
    enhanced,
    colback=lightgreen!60!white,
    colframe=green!25!gray,
    coltitle=black,
    boxrule=0.5pt,
    arc=0.5mm,
    top=2mm, bottom=2mm, left=2mm, right=2mm,
    fonttitle=\usefont{T1}{ppl}{b}{n},
    title=\model: Fundamentals Analyst Team -- Apple (AAPL),
]
\usefont{T1}{ppl}{m}{n}
\footnotesize
\begin{lstlisting}[style=agentbase]
[Fundamentals Analyst]:
================================== Report ==================================
### Comprehensive Fundamental Analysis Report for Apple Inc. (AAPL)
#### Company Profile
Apple Inc. is a prominent player in the Technology sector, known for its innovative products and significant market influence. Incorporated in 1980, Apple has a market capitalization of approximately $3.55 trillion USD, with 15,115.82 million shares outstanding. The company is listed on the NASDAQ under the ticker AAPL.

#### Financial Overview
- **52-Week Price Range**: $164.075 - $237.49
- **Price Returns**:
- 5-Day: 2.805%
- 13-Week: 3.7264%
- 26-Week: 23.6604%
- 52-Week: 24.0587%
- Year-to-Date: 22.0225%
- **Relative Performance**: Underperformed the S&P 500 by -7.6652% over the past year.

#### Profitability Metrics
- **Gross Margin**: 46.21%
- **Operating Margin**: 31.51%
- **Net Profit Margin**: 23.97%
- **Return on Equity (ROE)**: 164.59%
- **Return on Assets (ROA)**: 25.68%

#### Growth Metrics
- **EPS Growth**:
- 3-Year: 2.71%
- 5-Year: 15.41%
- Quarterly YoY: -34%
- **Revenue Growth**:
- 3-Year: 2.25%
- 5-Year: 8.49%
- Quarterly YoY: 6.07%
#### Liquidity and Solvency
- **Current Ratio**: 0.8673
- **Quick Ratio**: 0.826
- **Long-Term Debt to Equity**: 1.5057
- **Total Debt to Equity**: 1.8723

#### Valuation Metrics
- **P/E Ratio**: 37.793
- **P/B Ratio**: 62.2048
- **P/CF Ratio**: 30.0517

#### Cash Flow and Dividends
- **Cash Flow Per Share**: $7.1978
- **Dividend Yield**: 0.4287%
- **Payout Ratio**: 16.25%

#### Insider Sentiment
The insider sentiment for 2024 shows a predominantly negative trend, with several months exhibiting the most negative Monthly Share Purchase Ratio (MSPR) possible. Notably, September showed a dramatic shift to positive sentiment, suggesting a temporary bullish outlook.

#### Insider Transactions
Recent insider transactions include significant sales by executives:
- **Chris Kondo** sold 4,130 shares at $228.87.
- **Arthur D. Levinson** sold 150,000 shares over two transactions at prices around $224.68 to $228.66.
- **Katherine L. Adams** gifted 8,000 shares.

#### Analysis and Insights
- **Strengths**: Apple demonstrates strong profitability and growth potential, supported by high margins and efficient asset utilization.
- **Weaknesses**: Liquidity constraints and high leverage pose potential risks, especially in a rising interest rate environment.
- **Valuation**: The stock is trading at a premium, which may be justified by its strong brand and growth prospects, but caution is advised due to potential overvaluation.

#### Investment Considerations
- **Caution on Valuation**: Insider sales suggest potential concerns about stock valuation. Investors should consider reviewing their positions.
- **Monitor Insider Activity**: Continued monitoring of insider transactions is recommended to validate concerns about stock valuation or company performance.
- **Short-term Strategy**: Given the negative insider sentiment, short-term bearish strategies might be considered.

#### Conclusion
Apple Inc. remains a financially robust company with impressive profitability and growth metrics. However, potential investors should weigh the high valuation and leverage against the company's strong market position and growth potential. The insider sentiment and transactions suggest caution, particularly regarding stock valuation. Investors should consider these factors alongside broader market conditions and company performance when making investment decisions.
================================ Tool Calls ================================
Tool Calls:
  get_finnhub_company_profile (call_ScxiITX7Hf6yUuOr8g41a6WP)
 Call ID: call_ScxiITX7Hf6yUuOr8g41a6WP
  Args:
    ticker: AAPL
Tool Calls:
  get_finnhub_company_financials_history (call_jwgyq9XbocVWIMsOpagIArQO)
 Call ID: call_jwgyq9XbocVWIMsOpagIArQO
  Args:
    ticker: AAPL
    freq: quarterly
    end_date: 2024-11-19
  get_finnhub_basic_company_financials (call_o2JV2EjmMuTK3Z85debseS9C)
 Call ID: call_o2JV2EjmMuTK3Z85debseS9C
  Args:
    ticker: AAPL
  get_finnhub_company_insider_sentiment (call_DqhGKetFeNB3zWlHLJdrTE8H)
 Call ID: call_DqhGKetFeNB3zWlHLJdrTE8H
  Args:
    ticker: AAPL
    curr_date: 2024-11-19
  get_finnhub_company_insider_transactions (call_v78N7hIUA4CufM6EIgErEroZ)
 Call ID: call_v78N7hIUA4CufM6EIgErEroZ
  Args:
    ticker: AAPL
    curr_date: 2024-11-19
\end{lstlisting}
\end{tcolorbox}

% Researcher Team Bullish Opinion - Light Yellow
\begin{tcolorbox}[
    breakable,
    enhanced,
    colback=lightyellow!60!white,
    colframe=orange!20!gray,
    coltitle=black,
    boxrule=0.5pt,
    arc=0.5mm,
    top=2mm, bottom=2mm, left=2mm, right=2mm,
    fonttitle=\usefont{T1}{ppl}{b}{n},
    title=\model: Researcher Team Bullish Opinion -- Apple (AAPL),
]
\usefont{T1}{ppl}{m}{n}
\footnotesize
\begin{lstlisting}[style=agentbase]
[Bullish Researcher]
Bull Researcher: Alright, let's dive into why Apple is a solid investment opportunity right now, despite some of the bearish concerns that have been raised.

First off, let's talk about growth potential. Apple's expansion into the smart home market with a new AI-powered device is a significant move. This isn't just about keeping up with competitors like Amazon and Google; it's about leveraging their existing ecosystem to create a seamless user experience. Apple's ability to integrate new products into its ecosystem is a huge competitive advantage. This isn't just a new product; it's a strategic expansion that could drive significant revenue growth.

Now, I know the bear argument might point to challenges in China due to geopolitical tensions. But let's not forget that Apple has a history of navigating complex international markets successfully. Their strong brand loyalty and innovative product lineup give them a buffer against these challenges. Plus, the global demand for Apple's products remains robust, as evidenced by their impressive revenue growth metrics over the past five years.

Speaking of financials, Apple's profitability metrics are outstanding. With a gross margin of 46.21% and a net profit margin of 23.97%, Apple is operating at a level of efficiency that most companies can only dream of. Yes, the P/E ratio is high, but that's often the case with companies that have strong growth prospects and a dominant market position. The high valuation reflects investor confidence in Apple's future earnings potential.

The bear argument might also highlight insider sales as a red flag. However, insider sales can happen for a variety of reasons, not necessarily because insiders believe the stock is overvalued. It's crucial to look at the broader picture. The fact that major investment firms are increasing their holdings in Apple suggests that institutional investors see long-term value here.

Let's address the liquidity and leverage concerns. While Apple's current and quick ratios are below 1, indicating potential liquidity constraints, it's important to remember that Apple generates substantial cash flow. Their cash flow per share is strong, and they have a history of effectively managing their debt. This financial strength gives them the flexibility to invest in growth opportunities and weather economic uncertainties.

Finally, let's talk about market sentiment. Despite some fluctuations, the overall sentiment trend for Apple is predominantly positive. The spikes in positive sentiment, particularly on days with favorable news, indicate that the market is generally optimistic about Apple's prospects. This aligns with the technical indicators showing a bullish market sentiment, such as the upward trend in Bollinger Bands and MACD values.

In conclusion, while there are always risks to consider, Apple's strong growth potential, competitive advantages, and robust financial performance make it a compelling investment opportunity. The bearish concerns, while valid, are outweighed by the company's strengths and the positive market sentiment. Investing in Apple means betting on a company with a proven track record of innovation and success.
\end{lstlisting}
\end{tcolorbox}

% Researcher Team Bearish Opinion - Light Pink
\begin{tcolorbox}[
    breakable,
    enhanced,
    colback=lightpink!50!white,
    colframe=red!20!gray,
    coltitle=black,
    boxrule=0.5pt,
    arc=0.5mm,
    top=2mm, bottom=2mm, left=2mm, right=2mm,
    fonttitle=\usefont{T1}{ppl}{b}{n},
    title=\model: Researcher Team Bearish Opinion -- Apple (AAPL),
]
\usefont{T1}{ppl}{m}{n}
\footnotesize
\begin{lstlisting}[style=agentbase]
[Bearish Researcher]
Bear Researcher: Alright, let's dive into why investing in Apple right now might be riskier than it seems, despite the bullish enthusiasm.

First, let's tackle the growth potential argument. Yes, Apple is expanding into the smart home market with a new AI-powered device, but let's not overlook the intense competition from established players like Amazon and Google. These companies have been in the smart home space for years and have a significant head start. Apple's late entry could mean they face an uphill battle to capture market share, and there's no guarantee that their ecosystem will be enough to sway consumers away from competitors.

Now, about the geopolitical tensions with China. The bullish view suggests Apple can navigate these challenges due to brand loyalty and innovation. However, the reality is that geopolitical tensions can lead to sudden regulatory changes, tariffs, or even consumer boycotts, which could severely impact Apple's supply chain and sales in one of its largest markets. The risks here are not just theoretical; they're very real and could have substantial financial implications.

Regarding financials, while Apple's profitability metrics are indeed strong, the high PE ratio is a red flag. It suggests that the stock is priced for perfection, leaving little room for error. Any misstep or negative news could lead to a significant correction. The bullish argument glosses over this by attributing it to investor confidence, but high valuations can also indicate overvaluation, which is a risk in itself.

Insider sales are another point of concern. While it's true that insiders sell for various reasons, the pattern of significant sales by top executives could indicate a lack of confidence in the stock's current valuation. This isn't something to dismiss lightly, especially when combined with the negative insider sentiment observed throughout the year.

On liquidity and leverage, the bullish argument mentions Apple's strong cash flow, but the current and quick ratios below 1 are concerning. This indicates potential liquidity issues, which could be problematic if economic conditions worsen or if Apple needs to make significant investments to stay competitive. High leverage also means increased risk, especially in a rising interest rate environment where debt servicing costs could escalate.

Finally, let's address market sentiment. While there have been spikes in positive sentiment, the overall picture is mixed, with notable negative sentiment days. This volatility in sentiment suggests that the market is uncertain about Apple's future, which aligns with the technical indicators showing potential for price corrections and ongoing volatility.

In conclusion, while Apple is undoubtedly a strong company with a history of success, the current investment landscape is fraught with risks. High valuation, geopolitical tensions, insider sales, and liquidity concerns all point to potential vulnerabilities. The bullish arguments, while highlighting Apple's strengths, don't fully account for these risks. As a bear, I believe it's crucial to approach Apple with caution, considering the potential for downside in the current market environment.
\end{lstlisting}
\end{tcolorbox}

% Risk Management Team Aggressive Agent - Light Purple
\begin{tcolorbox}[
    breakable,
    enhanced,
    colback=lightpurple!40!white,
    colframe=violet!25!gray,
    coltitle=black,
    boxrule=0.5pt,
    arc=0.5mm,
    top=2mm, bottom=2mm, left=2mm, right=2mm,
    fonttitle=\usefont{T1}{ppl}{b}{n},
    title=\model: Risk Management Team Aggressive Agent -- Apple (AAPL),
]
\usefont{T1}{ppl}{m}{n}
\footnotesize
\begin{lstlisting}[style=agentbase]
[Risky Analyst]
Risky Analyst: Let's dive into why a high-reward, high-risk approach with Apple Inc. (AAPL) is the optimal strategy right now, especially when considering the current market dynamics and sentiment.

First, let's address the conservative analyst's likely concerns about the potential for price corrections due to overbought conditions and high valuation. While it's true that the RSI and CCI indicate overbought conditions, these are signs of strong upward momentum. The market is clearly bullish, as evidenced by the consistent upward trend in Bollinger Bands and MACD values. This bullish sentiment is further supported by the positive social media sentiment peaks on November 15 and 11, suggesting favorable developments that could drive the stock higher. By focusing too much on potential corrections, the conservative view might miss the opportunity to capitalize on this momentum.

Now, regarding the neutral analyst's probable stance on the weakening trend strength indicated by the ADX and the stable supertrend suggesting consolidation, these are precisely the conditions that set the stage for potential breakouts. The market's current consolidation phase, coupled with the heightened volatility indicated by the ATR, presents a ripe environment for significant upward movements. The neutral perspective might overlook the potential for these breakouts, which could lead to substantial gains for those willing to take the risk.

Furthermore, the global economic and political environment, while presenting challenges such as US-China tensions, also offers opportunities. The return of Donald Trump and potential deregulation could spur innovation and spending in the tech sector, benefiting companies like Apple. The expansion into the smart home market with AI-powered devices positions Apple to capture new growth areas, which the conservative and neutral views might underestimate.

From a fundamental standpoint, Apple's strong profitability metrics, such as a 46.21% gross margin and a 164.59% ROE, highlight its ability to generate significant returns. While concerns about liquidity and leverage exist, Apple's robust cash flow and market position provide a buffer against these risks. The conservative focus on insider sales and high valuation might miss the broader picture of Apple's strategic positioning and growth potential.

In conclusion, while there are risks associated with investing in Apple at this juncture, the potential rewards far outweigh them. The market's bullish sentiment, coupled with Apple's strategic initiatives and strong financials, presents a compelling case for a high-reward approach. By embracing this strategy, traders can position themselves to outpace market norms and capitalize on the opportunities that others might overlook due to excessive caution.
Safe Analyst: Let's take a closer look at the arguments presented by the Risky Analyst and why a conservative approach is more prudent in this situation.
\end{lstlisting}
\end{tcolorbox}

% Risk Management Team Conservative Agent - Light Orange
\begin{tcolorbox}[
    breakable,
    enhanced,
    colback=lightorange!60!white,
    colframe=orange!30!gray,
    coltitle=black,
    boxrule=0.5pt,
    arc=0.5mm,
    top=2mm, bottom=2mm, left=2mm, right=2mm,
    fonttitle=\usefont{T1}{ppl}{b}{n},
    title=\model: Risk Management Team Conservative Agent -- Apple (AAPL),
]
\usefont{T1}{ppl}{m}{n}
\footnotesize
\begin{lstlisting}[style=agentbase]
[Safe Analyst]
Firstly, the Risky Analyst highlights the overbought conditions indicated by the RSI and CCI as signs of strong upward momentum. However, these indicators also suggest that the stock is potentially overvalued and ripe for a correction. Ignoring these signals could lead to significant losses if the market corrects, which is a real possibility given the high volatility indicated by the ATR. The conservative approach would be to wait for these indicators to stabilize before making any aggressive moves, thereby minimizing the risk of buying at a peak.

Regarding the potential for breakouts during the consolidation phase, as suggested by the ADX and supertrend indicators, it's important to remember that consolidation can also lead to breakdowns, not just breakouts. The heightened volatility and fluctuating volume further increase the uncertainty. A cautious strategy would involve waiting for clearer signals of a breakout before committing significant resources, thus protecting the firm's assets from unnecessary exposure to risk.

The Risky Analyst also mentions the potential benefits of deregulation and Apple's expansion into new markets. While these are positive developments, they are speculative and contingent on numerous external factors, such as geopolitical tensions and economic policies, which are currently unpredictable. The ongoing US-China tensions, for instance, pose a significant risk to Apple's operations and supply chain, which could negatively impact its stock performance. A conservative approach would involve closely monitoring these developments and adjusting strategies accordingly, rather than making premature investments based on uncertain outcomes.

From a fundamental perspective, while Apple's profitability metrics are indeed strong, the high valuation and insider sales are red flags that cannot be ignored. The P/E and P/B ratios suggest that the stock is trading at a premium, which could lead to a sharp decline if market sentiment shifts. Insider sales further indicate that those closest to the company might believe the stock is overvalued. A conservative strategy would involve reassessing the valuation and waiting for a more favorable entry point, ensuring that investments are made at a sustainable price level.

In conclusion, while the potential for high rewards exists, the associated risks are substantial and could jeopardize the firm's assets. A conservative approach prioritizes stability and long-term growth by carefully evaluating market conditions and avoiding unnecessary exposure to volatility. By focusing on risk mitigation and asset protection, we can ensure steady, reliable growth without succumbing to the allure of short-term gains that may not be sustainable.
Neutral Analyst: Alright, let's dive into this discussion and find a balanced perspective between the Risky and Safe Analysts' viewpoints on Apple Inc. (AAPL).

\end{lstlisting}
\end{tcolorbox}

% Risk Management Team Neutral Agent - Light Red
\begin{tcolorbox}[
    breakable,
    enhanced,
    colback=lightred!50!white,
    colframe=red!25!gray,
    coltitle=black,
    boxrule=0.5pt,
    arc=0.5mm,
    top=2mm, bottom=2mm, left=2mm, right=2mm,
    fonttitle=\usefont{T1}{ppl}{b}{n},
    title=\model: Risk Management Team Neutral Agent -- Apple (AAPL),
]
\usefont{T1}{ppl}{m}{n}
\footnotesize
\begin{lstlisting}[style=agentbase]
[Neutral Analyst]
Starting with the Risky Analyst's perspective, there's a strong emphasis on the bullish momentum and potential for high rewards. It's true that indicators like the RSI and MACD suggest upward momentum, and the positive sentiment peaks could signal favorable developments. However, the Risky Analyst might be underestimating the potential for a market correction. The overbought conditions and high volatility indicated by the ATR suggest that the market could swing either way. While the potential for breakouts exists, it's crucial to remember that these conditions can also lead to significant downturns. A more balanced approach would involve setting stop-loss orders or considering options strategies to hedge against potential losses while still participating in the upside.

On the other hand, the Safe Analyst is right to point out the risks associated with overvaluation and insider sales. The high P/E and P/B ratios do suggest that Apple might be trading at a premium, and insider sales could be a red flag. However, the Safe Analyst might be overly cautious by suggesting waiting for a more favorable entry point. Given Apple's strong profitability metrics and strategic initiatives, there could be missed opportunities if one waits too long. Instead, a moderate approach could involve gradually building a position in Apple, allowing for dollar-cost averaging to mitigate the risk of buying at a peak.

Regarding the geopolitical and economic factors, both analysts have valid points. The Risky Analyst sees potential in deregulation and Apple's expansion into new markets, while the Safe Analyst rightly highlights the unpredictability of these factors. A balanced strategy would involve closely monitoring these developments and being ready to adjust positions as new information becomes available. Diversifying investments across different sectors or geographies could also help mitigate the risks associated with geopolitical tensions.

In conclusion, a moderate risk strategy would involve a careful balance of taking advantage of Apple's growth potential while safeguarding against extreme volatility. This could include using a mix of technical analysis to time entries and exits, fundamental analysis to assess long-term value, and diversification to spread risk. By adopting this balanced approach, investors can aim for steady growth without exposing themselves to unnecessary risks.
\end{lstlisting}
\end{tcolorbox}

% Fund Manager - Light Lilac
\begin{tcolorbox}[
    breakable,
    enhanced,
    colback=lightlilac!50!white,
    colframe=violet!20!gray,
    coltitle=black,
    boxrule=0.5pt,
    arc=0.5mm,
    top=2mm, bottom=2mm, left=2mm, right=2mm,
    fonttitle=\usefont{T1}{ppl}{b}{n},
    title=\model: Fund Manager -- Apple (AAPL),
]
\usefont{T1}{ppl}{m}{n}
\footnotesize
\begin{lstlisting}[style=agentbase]
[Manager]
Based on the investment plan provided and the analysis of both bullish and bearish perspectives, I concur with the recommendation to **Buy** Apple (AAPL) shares. Here is a summary of the rationale:

1. **Strong Fundamentals and Growth Prospects:** Apple's consistent earnings growth and robust market capitalization highlight its financial strength and potential for continued expansion. The company's innovation, particularly in AI and other cutting-edge technologies, positions it well for future growth.

2. **Innovation and Market Leadership:** Apple's commitment to innovation and maintaining a competitive edge through its ecosystem strengthens its market position. This is crucial in sustaining long-term growth and profitability.

3. **Resilience to External Risks:** Despite concerns about geopolitical tensions and financial risks, Apple's strategic partnerships and global supply chain management mitigate these challenges effectively.

4. **Positive Market Sentiment:** The overall positive sentiment towards Apple, despite some volatility, indicates strong investor confidence in its long-term prospects.

While the concerns about high valuation ratios and insider selling are valid, they are outweighed by Apple's strong fundamentals and growth potential. The high P/E ratio can be justified by Apple's market leadership and sustained growth trajectory. Additionally, the proposed risk management strategies, such as setting stop-loss orders and maintaining a diversified portfolio, provide a prudent approach to mitigating potential risks.

Therefore, considering the comprehensive analysis and the investment plan's emphasis on long-term growth and resilience, I recommend a **Buy** for Apple shares.

\end{lstlisting}
\end{tcolorbox}

\end{document}